\documentclass[11pt]{article}
\usepackage[margin=2cm]{geometry}
\usepackage{color} 

\usepackage{amsmath,bbm,latexsym,amssymb}
\usepackage{mathtools}
\usepackage{graphicx}
\usepackage{rotating}
\usepackage{color} 
\usepackage{multirow} 
\usepackage[T1]{fontenc} 
\usepackage{hyperref}
\usepackage{lmodern}
\makeatletter

\usepackage{booktabs}
\usepackage{mathrsfs}   
\usepackage{slashed}     
\usepackage{url}
\usepackage{units}
\usepackage{wasysym }
\usepackage{comment}

\usepackage{float}
\newcommand{\bea}{\begin{eqnarray}}
\newcommand{\eea}{\end{eqnarray}}

\newcommand{\be}{\begin{equation}}
\newcommand{\ee}{\end{equation}}
\def\beq{\begin{equation}}
\def\eeq{\end{equation}}
\newcommand{\ba}{\begin{eqnarray}}
\newcommand{\ea}{\end{eqnarray}}

 \def\ifmath#1{\relax\ifmmode #1\else $#1$\fi}

\makeatother

\def\vb#1{\vbox to #1 pt{}}

\def\vb#1{\vbox to #1 pt{}}

\def\gsim{\raise0.3ex\hbox{$\;>$\kern-0.75em\raise-1.1ex\hbox{$\sim\;$}}}
\def\lsim{\raise0.3ex\hbox{$\;<$\kern-0.75em\raise-1.1ex\hbox{$\sim\;$}}}

\allowdisplaybreaks

\usepackage{dsfont}

\begin{document}

\title{\begin{flushright}
\vspace{-3mm}
\small{CFTP/23-001}\\[1mm]
\end{flushright}
\textbf{A viable $A_4$ 3HDM theory of quark mass matrices}}

\author{
Iris Br\'{e}e\footnote{ iris.bree.silva@tecnico.ulisboa.pt},
S\'{e}rgio Carr\^{o}lo \footnote{sergio.carrolo@tecnico.ulisboa.pt},
Jorge C. Rom\~{a}o \footnote{jorge.romao@tecnico.ulisboa.pt},
Jo\~{a}o P. Silva \footnote{jpsilva@cftp.ist.utl.pt},
\\
CFTP, Departamento de F\'{\i}sica,\\
Instituto Superior T\'{e}cnico, Universidade de Lisboa, \\
Avenida Rovisco Pais 1, 1049 Lisboa, Portugal \\[5pt]
}

\maketitle

\begin{abstract}
It is known that a three Higgs doublet model (3HDM)
symmetric under an exact $A_4$ symmetry is not compatible
with nonzero quark masses and/or non-block-diagonal CKM matrix.
We show that a 3HDM with softly broken $A_4$ terms in the
scalar potential does allow for a fit of quark mass matrices.
Moreover, the result is consistent with $m_h=125\textrm{GeV}$ and
the $h \rightarrow WW,ZZ$ signal. We also checked numerically
that, for each point that passes all the constraints, the minimum is
a global minimum of the potential.
\end{abstract}

\section{Introduction}

The observation in 2012 of a scalar particle with 125GeV by the ATLAS
and CMS collaborations \cite{ATLAS:2012yve,CMS:2012qbp} has
incentivized experimental searches for beyond the Standard Model (SM)
particles at the LHC. On par with these experimental endeavors,
theoretical efforts in the search for extra scalar particles have been
strengthened since this discovery. A promising framework is found in
N-Higgs doublet models (NHDM).  

Such models have many free parameters, which are often curtailed by imposing
some discrete family symmetry.
Here, we focus on the implementation of $A_4$ in a three Higgs doublet model
(3HDM).
The $A_4$ group is the group of even permutations on 4 elements. It is
the smallest discrete group to contain a three-dimensional irreducible representation (irrep),
which is ideal for describing the three families of quarks with a
minimal number of independent Yukawa couplings.
Thus, NHDM supplemented by the $A_4$ discrete symmetry has long been of interest
in flavour physics research.

A number of early articles include:
\cite{Ma:2001dn},
mainly devoted to the leptonic sector and where the solution to the quark sector
is briefly mentioned to include a fourth Higgs doublet and all quark fields in singlets
(which is effectively the same as the Standard Model quark sector);
\cite{Ma:2002yp},
where $A_4$ is broken by dimension four Yukawa couplings,
which, upon renormalization, will affect the scalar potential
\cite{Ma:2006sk},
which requires three Higgs doublets in the down-type quark sector and a further
two in the up-type quark sector, consisting of a 5HDM;
and \cite{Ma:2006wm},
which is devoted to the leptonic sector, but has the interesting side query that
it might be possible to recover a realistic CKM matrix through soft-breaking
of $A_4$.

Quark mass matrices in the context of a 3HDM with Higgs doublets in the triplet
representation of $A_4$ were studied in \cite{Lavoura:2007dw} and \cite{Morisi:2009sc},
with the vacuum expectation value (vev) structure $(e^{i\alpha}, e^{- i \alpha}, r)$,
where $\alpha$ and $r$ are real constants.
This vacuum solution was also included in the original
study of the $A_4$-3HDM vaccua in Ref.~\cite{deAdelhartToorop:2010jxh}.
Unfortunately,
Degee, Ivanov and Keus \cite{Degee:2012sk} proved in 2013 that such a vacuum can
\textit{never} be the global minimum of the $A_4$ symmetric 3HDM.
In this beautiful paper,
geometric techniques were used in order to identify all possible global minima
(thus, all possible viable vacua) of the $A_4$ symmetric 3HDM.
Immediately thereafter,
those minima were used to show that all assignments
of the quark fields into irreps of $A_4$,
when combined with the possible vevs for the exact
$A_4$ potential, yield vanishing quark masses and/or a
CP conserving CKM matrix, both of which are forbidden by
experiment.
This is in fact a consequence of a much broader theorem,
proved in \cite{Leurer:1992wg,GonzalezFelipe:2014mcf}:
given any flavour symmetry group,
one can obtain a physical CKM mixing matrix and, simultaneously,
non-degenerate and non-zero quark masses only if the
vevs of the Higgs fields break completely the full flavour group.
The idea is that a symmetry will reduce the number of redundant
Yukawa couplings present in the SM, and it might even predict
relations among observables which turn out to be consistent
with experiment.

When studying in detail the extensions of $A_4$ to the quark sector
found by Ref.~\cite{GonzalezFelipe:2013xok}, we noticed that, in some
of them, if it weren't for the particular form of the vevs allowed by
the exact $A_4$ 3HDM potential, the Yukawa
matrices could allow for massive quarks, and for a realistic CKM
matrix. Since the $A_4$ symmetric potential doesn't allow for
minima other than those shown in \cite{Degee:2012sk}, here we
consider the case where the $A_4$ symmetry is softly broken by the
addition of quadratic terms to the potential. 
Such terms do not spoil the theory's renormalizability, but break the
$A_4$ symmetry.

Our article is organized as follows. We define the notation for the
scalar potential in Sec.~\ref{subsec:pot},
discuss the Yukawa Lagrangian and the form of the possible mass matrices in
Sec.~\ref{subsec:Yuk}, giving all the
expressions needed for the fit in Sec.~\ref{subsec:Yuk2}.
In Sec.~\ref{sec:fitting} we present our fit to the
quarks mass matrices,
while in Sec.~\ref{sec:viable}
we discuss the viability of the vacuum found in the fit in terms of the scalar potential.
Sec.~\ref{sec:theorconstraints} is devoted to the implementation of the theoretical constraints to
be imposed, and in Sec.~\ref{sec:LHC} we briefly discuss the constraints coming
from the LHC.
The results and conclusions are presented in Sec.~\ref{sec:results}
and \ref{sec:concl}, respectively.
The Appendices contain some additional expressions that are needed for the fits.

\section{Parameterization for the softly-broken $A_4$ 3HDM}

\subsection{\label{subsec:pot}Potential and candidates for local minimum}

The softly-broken potential of the 3HDM with an $A_4$ symmetry is given by
\begin{equation}
    V_H=V_{4,\, A_4}+M^2_{ij}\left( \phi_i^{\dagger} \phi_j \right) \, ,
\end{equation}
where $V_{4,\, A_4}$ is the quartic potential for the $A_4$ symmetric
three Higgs doublet model (3HDM),
which is, in the notation of \cite{Degee:2012sk},

\begin{align}
V_{4,\, A_4} =& 
\frac{\Lambda_0}{3}  \left(\phi_1 ^ \dagger \phi_1
+ \phi_2 ^ \dagger \phi_2 + \phi_3 ^ \dagger \phi_3\right) ^ 2
\nonumber\\*[1mm]
&
+ \Lambda_1 \left[ 
\left( \textrm{Re}\left\{\phi_1^\dagger\phi_2\right\} \right)^2
+ \left( \textrm{Re}\left\{\phi_2^\dagger\phi_3\right\} \right)^2
+ \left( \textrm{Re}\left\{\phi_3^\dagger\phi_1\right\} \right)^2
\right]
\nonumber\\*[1mm]
& 
+ \Lambda_2 \left[ 
\left( \textrm{Im}\left\{\phi_1^\dagger\phi_2\right\} \right)^2
+ \left( \textrm{Im}\left\{\phi_2^\dagger\phi_3\right\} \right)^2
+ \left( \textrm{Im}\left\{\phi_3^\dagger\phi_1\right\} \right)^2
\right]
\nonumber\\*[1mm]
&
+ \frac{\Lambda_3}{3} \left[ (\phi_1 ^ \dagger \phi_1) ^ 2
+ (\phi_2 ^ \dagger \phi_2) ^ 2 + (\phi_3 ^ \dagger \phi_3) ^ 2 
- (\phi_1^\dagger \phi_1) (\phi_2^\dagger \phi_2)
- (\phi_2^\dagger \phi_2) (\phi_3^\dagger \phi_3)
- (\phi_3^\dagger \phi_3) (\phi_1^\dagger \phi_1)\right]
\nonumber\\*[1mm]
&
+ \Lambda_4 \left[
\textrm{Re}\left\{\phi_1^\dagger\phi_2\right\}\,
	\textrm{Im}\left\{\phi_1^\dagger\phi_2\right\}
+  \textrm{Re}\left\{\phi_2^\dagger\phi_3\right\}\,
	\textrm{Im}\left\{\phi_2^\dagger\phi_3\right\}
+ \textrm{Re}\left\{\phi_3^\dagger\phi_1\right\}\,
	\textrm{Im}\left\{\phi_3^\dagger\phi_1\right\}
\right] \,.
\label{V_A4}
\end{align}
The matrix $M^2_{ij}$ is a general hermitian matrix, which can be
parameterized by 
\begin{equation}
\label{eq:soft}
(M^2_{ij}) =
\begin{pmatrix}
m^2_{11} & m^2_{12} e^{i\theta_{12}} & m^2_{13} e^{i\theta_{13}}\\
m^2_{12} e^{-i\theta_{12}} & m^2_{22} & m^2_{23} e^{i\theta_{23}}\\
m^2_{13} e^{-i\theta_{13}} & m^2_{23} e^{-i\theta_{23}} & m^2_{33}
\end{pmatrix}\, ,
\end{equation}
where $m^2_{ij}$ are real parameters with the dimension
of mass squared.\footnote{In the quadratic terms,
the combination
$- \frac{M_0}{\sqrt{3}} \left(\phi_1 ^ \dagger \phi_1
+ \phi_2 ^ \dagger \phi_2 + \phi_3 ^ \dagger \phi_3\right)$
is also invariant under $A_4$.
But, since we are keeping all soft-breaking terms,
we find the notation in \eqref{eq:soft} more convenient.
}

Additionally, in the notation of \cite{Bento:2022vsb}, the exact $A_4$
potential can be written as 
\begin{align}
    V_{A_4} =& \frac{r_1 + 2 r_4}{3} \left[ (\phi_1^\dagger \phi_1) 
    + (\phi_2^\dagger \phi_2) + (\phi_3^\dagger \phi_3) \right]^2
    + \frac{2 (r_1 - r_4)}{3} \left[ (\phi_1^\dagger \phi_1)^2 
    + (\phi_2^\dagger \phi_2)^2  \right.
    \nonumber \\[2mm]
    &\left. + (\phi_3^\dagger \phi_3)^2
    - (\phi_1^\dagger \phi_1) (\phi_2^\dagger \phi_2)
    - (\phi_2^\dagger \phi_2) (\phi_3^\dagger \phi_3)
    - (\phi_3^\dagger \phi_3) (\phi_1^\dagger \phi_1)\right]
    \nonumber \\[2mm]
    &+ 2 r_7 \left( |\phi_1^\dagger \phi_2|^2 + |\phi_2^\dagger \phi_3|^2 
    + |\phi_3^\dagger \phi_1|^2 \right)
    \nonumber \\[2mm]
    & + \Big[
	c_3 \left[ (\phi_1^\dagger \phi_2)^2 + (\phi_2^\dagger \phi_3)^2 
    + (\phi_3^\dagger \phi_1)^2 \right]
    + h.c. \Big] \, .
\label{V4_A4_one}
\end{align}
The relation between the two notations is
\begin{align}
  &&r_1=\frac{1}{3}(\Lambda_0+\Lambda_3) \, , &&
  r_4=\frac{1}{6}(2 \Lambda_0-\Lambda_3) \, , &&
  r_7=\frac{1}{4}(\Lambda_1+\Lambda_2) \, , \nonumber\\
  &&\text{Re}(c_3)= \frac{1}{4}(\Lambda_1-\Lambda_2) \, ,&&
  \text{Im}(c_3)= -\frac{1}{4} \Lambda_4 \, . &&
\end{align}

We consider that the scalar fields can take complex vacuum expectation
values (vevs),  to be
determined later.
Thus, we write,
\begin{equation}
  \label{eq:2}
  \phi_i=
  \begin{bmatrix}
    \varphi^+_i\\
    \frac{|v_i| e^{i \rho_i}}{\sqrt{2}} + \frac{1}{\sqrt{2}} \left( x_i +
        i x_{i+3} \right) 
  \end{bmatrix} .
\end{equation}
Because CP is spontaneously violated, the unrotated neutral fields have no
definite CP, and for convenience we label them $x_i,i=1,\ldots,6$. We
can also use the gauge freedom to absorb one of the phases in the
vevs, that we choose to be $\rho_1$. Therefore we have 
the vector of  vevs defined as
\begin{equation}
\vec{v} = (|v_1|,|v_2| e^{i\rho_2},|v_3| e^{i\rho_3})\, .
\end{equation}
This vev contributes with four free parameters to our model, because
one of the parameters is constrained by the mass of the gauge bosons to
match the observed SM values,
\begin{equation}
    |v_1|^2+ |v_2|^2+ |v_3|^2 \equiv v^2 \simeq (246\textrm{GeV})^2.
\end{equation}
The vev can also be parameterized as
\begin{equation}
    \vec{v} =
v \left(\cos(\beta_1)\cos(\beta_2), \cos(\beta_2)\sin(\beta_1) e^{ip_2},
\sin(\beta_2)e^{ip_3}\right).
\label{eq:vev_par}
\end{equation}

Of the quantities arising out of the scalar potential,
the vevs are the only relevant to the quark mass matrices.
This leads many authors to just proclaim some vevs, without checking whether
they can indeed be the global minima of a realistic Higgs potential.
We will perform this crucial verification below,
in Section~\ref{sec:viable}.

\subsection{\label{subsec:Yuk}Yukawa Lagrangian}

As in Refs.~\cite{Lavoura:2007dw,GonzalezFelipe:2013xok},
we consider
that the Higgs doublets 
are in the $\textbf{3}$ of $A_4$ as well as the three left-handed
 $SU(2)$ doublets $Q_{Lj}$ of hypercharge 1/6. There are three right-handed
 $SU(2)$ singlets $n_{R,j}$ of hypercharge $-1/3$ and  three right-handed
 $SU(2)$ singlets $p_{R,j}$ of hypercharge $2/3$. Our assignments for
 the singlets are as follows
 \begin{align}
   n_{R1}, p_{R1}\to \textbf{1}, \quad
   n_{R2}, p_{R2} \to \textbf{1}^\prime, \quad
   n_{R3}, p_{R3}  \to \textbf{1}^{\prime\prime} \
   \text{   of   } A_4\, .
 \end{align}
Then, the $A_4$ transformations on the fields
are generated by \cite{Lavoura:2007dw,GonzalezFelipe:2013xok}
\begin{align}
  \label{eq:1}
  T:
  \left\{
  \begin{matrix*}[l]
    \phi_1\to\phi_2\to\phi_3\to\phi_1,\\[+1mm]
    Q_{L1}\to Q_{L2}\to Q_{L3}\to Q_{L1},\\[+1mm]
    n_{R1}\to n_{R1},  n_{R2}\to \omega n_{R2}, n_{R3}\to \omega^2
    n_{R3},\\[+1mm]
    p_{R1}\to n_{R1},  p_{R2}\to \omega p_{R2}, p_{R3}\to \omega^2 p_{R3},
  \end{matrix*}
  \right.
\end{align}
and 
\begin{align}
  \label{eq:14}
  S:
  \left\{
  \begin{matrix*}[l]
    \phi_1\to \phi_1, \phi_2\to -\phi_2, \phi_3\to -\phi_3,\\[+1mm]
    Q_{L1}\to Q_{L1}, Q_{L2}\to -Q_{L2}, Q_{L3}\to -Q_{L3}.\\[+1mm]
  \end{matrix*}
  \right.
\end{align}
One can easily verify that the scalar potential in Eq.~(\ref{V4_A4_one})
is invariant under the previous transformations. Now we write the
$A_4$ invariant Yukawa Lagrangian for quarks. We have
\begin{align}
  \label{eq:16}
-\mathcal{L}_{\rm Yukawa}=&
\sqrt{2}\, \hat{a}\left(
\overline{Q_{L1}}\phi_1+ \overline{Q_{L2}}\phi_2+\overline{Q_{L3}}\phi_3
\right) n_{R1}\nonumber\\[+1mm]
&+\sqrt{2}\, \hat{b}\left(
  \overline{Q_{L1}}\phi_1+ \omega\, \overline{Q_{L2}}\phi_2
  +\omega^2\, \overline{Q_{L3}}\phi_3
\right) n_{R2}\nonumber\\[+1mm]
&+\sqrt{2}\, \hat{c}\left(
  \overline{Q_{L1}}\phi_1+ \omega^2\,\overline{Q_{L2}}\phi_2
  +\omega\, \overline{Q_{L3}}\phi_3
\right) n_{R3}\nonumber\\[+1mm]
&+\sqrt{2}\, \hat{a}'\left(
\overline{Q_{L1}}\tilde{\phi}_1+ \overline{Q_{L2}}\tilde{\phi}_2+\overline{Q_{L3}}\tilde{\phi}_3
\right) p_{R1}\nonumber\\[+1mm]
&+\sqrt{2}\, \hat{b}'\left(
  \overline{Q_{L1}}\tilde{\phi}_1+ \omega\, \overline{Q_{L2}}\tilde{\phi}_2
  +\omega^2\, \overline{Q_{L3}}\tilde{\phi}_3
\right) p_{R2}\nonumber\\[+1mm]
&+\sqrt{2}\, \hat{c}'\left(
  \overline{Q_{L1}}\tilde{\phi}_1+ \omega^2\,\overline{Q_{L2}}\tilde{\phi}_2
  +\omega\, \overline{Q_{L3}}\tilde{\phi}_3
\right) p_{R3} + \text{h.c.} ,
\end{align}
where, as usual,
\begin{equation}
  \label{eq:17}
  \tilde{\phi}_j \equiv i\, \sigma_2 \phi^*_j\, ,
\end{equation}
and we define
\begin{equation}
  \label{eq:18}
  \hat{a}=a e^{i\, \alpha},\
  \hat{b}=b e^{i\, \beta},\
  \hat{c}=c e^{i\, \gamma},\
  \hat{a}'=a' e^{i\, \alpha'},\
  \hat{b}'=b' e^{i\, \beta'},\
  \hat{c}'=c' e^{i\, \gamma'}\, ,
\end{equation}
where $a,b,c,a',b',c'$ are real and positive. This choice of invariant
Lagrangian corresponds to the case I identified in
Ref.~\cite{GonzalezFelipe:2013xok} (see the next section).

\subsection{\label{subsec:Yuk2}Yukawa matrices, masses and CKM}

We aim to fit six quark masses and four CKM matrix elements to the currently
accepted SM values for these observables.
Therefore, we're interested in softly-broken $A_4$ symmetric models
with up to ten parameters.
Ref.~\cite{GonzalezFelipe:2013xok} has studied all of the possible extensions
of $A_4$ to the fermion sector.
Using their results, we can check which of them can accommodate
non-vanishing quark masses, CKM mixing angles and
CP violation by considering a general vev $\vec{v}$.
We take the Jarlskog invariant as a measure of
CP violation \cite{Jarlskog:1985ht}.
Out
of all possibilities, we are left with five of them, which we list in
Table~\ref{tb:CasesA4}. There, $A$ are real constants, $\Omega$ are constants
in the $[0, 2\pi[$ interval, $\omega=e^{i \frac{2 \pi}{3}}$
($\omega^3=1$) and $^T$ is the transpose of the matrix.

\begin{table}[H]
\centering
	\begin{tabular}{|c|c|c|}
	\hline
	Case & $M_d$ & $M_u$ \\
	\hline
\vb{30}	I & $\begin{pmatrix}
		a e^{i\alpha} v_1 & be^{i\beta}v_1 & ce^{i\gamma}v_1 \\
		a e^{i\alpha} v_2 & \omega be^{i\beta}v_2 & \omega^2 ce^{i\gamma}v_2 \\
		a e^{i\alpha} v_3 & \omega^2 be^{i\beta}v_3 & \omega ce^{i\gamma}v_3 \\
	\end{pmatrix}$ 	& $\begin{pmatrix}
		A \rightarrow A', & A\in \{a,b,c\} \\
		\Omega \rightarrow \Omega', & \Omega \in \{\alpha,\beta,\gamma\} \\
		v_i \rightarrow v_i^*, & i \in \{1,2,3\}
	\end{pmatrix}$	\\[+6.5mm]
	\hline
\vb{14}	II & $\text{I}_d^T$ & $\text{I}_u^T$ \\[+1mm]
	\hline 
\vb{30}	III & $\begin{pmatrix}
		0 & (ae^{i\alpha} - b e^{i\beta})v_3 & (ae^{i\alpha} + b e^{i\beta})v_2 \\
		(ae^{i\alpha} + b e^{i\beta})v_3 & 0 & (ae^{i\alpha} - b e^{i\beta})v_1 \\
		(ae^{i\alpha} - b e^{i\beta})v_2 & (ae^{i\alpha} + b e^{i\beta})v_1 & 0 
	\end{pmatrix}$ & $\begin{pmatrix}
		A \rightarrow A', & A\in \{a,b\} \\
		\Omega \rightarrow \Omega', & \Omega \in \{\alpha,\beta\} \\
		v_i \rightarrow v_i^*, & i \in \{1,2,3\}
	\end{pmatrix}$\\[+6.5mm]
	\hline 
\vb{14}	IV & $\text{I}_d$ & $\text{III}_u$ \\[+1mm]
	\hline
\vb{14}	V & $\text{III}_d$ & $\text{I}_u$ \\
	\hline
	\end{tabular}
\caption{Extensions of $A_4$ to the Yukawa sector with non-vanishing
  determinant, and non-zero $J$ for general, complex valued, vevs
  $(v_1,v_2,v_3)$. In the Table, $\text{I}_d$ stands for the matrix
  $M_d$ for case $\text{I}$ and similarly for the other entries.}
\label{tb:CasesA4}
\end{table}

\noindent
In the table above, we have used the convention where the quarks' mass
terms are written as 

\begin{equation}
	-\mathcal{L}_{\text{Yukawa}} \supset \overline{n_L} M_d n_R +
        \overline{p_L} M_u p_R + \text{h.c.}\, , 
\end{equation}
where h.c. stands for the hermitian conjugate.

In the Yukawa sector, there are ten observables, six masses, three
mixing angles and one Jarlskog invariant, therefore, we would prefer
to look for a case with ten parameters, or less. All possible neutral
vevs of the 3HDM are consistent with the parameterization in
Eq.~\eqref{eq:vev_par}, which consists of four free parameters that we can
fit; two angles, and two phases. Looking at the cases in
Table~\ref{tb:CasesA4}, we will see that it is possible to reduce the number
of free parameters by performing both basis transformations to
right-handed quarks and global $U(1)_Y$ rephasings, both of which have no
effect on the physical predictions of the theory.

For case I, the down quark mass matrices read

\begin{align}
\label{MdMu} M_d =
\begin{pmatrix}
		a e^{i\alpha} v_1 & be^{i\beta}v_1 & ce^{i\gamma}v_1
		\\ a e^{i\alpha} v_2 & \omega be^{i\beta}v_2 &
		\omega^2 ce^{i\gamma}v_2 \\ a e^{i\alpha} v_3 &
		\omega^2 be^{i\beta}v_3 & \omega ce^{i\gamma}v_3 \\
	\end{pmatrix}\, = D_v W D_a D_{\alpha} ,
\end{align}
where (remember that the $v_i$ are complex)
\begin{equation} D_{v}= \textrm{diag}(v_1, v_2, v_3)\, , \,\,
D_{a} = \textrm{diag}(a,b,c)\, , \,\, D_{\alpha} = \textrm{diag}(e^{i
\alpha},e^{i \beta},e^{i \gamma})\, , \,\, W =
\begin{pmatrix} 1 & 1 & 1\\ 1 & \omega & \omega^2 \\ 1 & \omega^2 &
\omega
\end{pmatrix}\, .
\label{eq:11}
\end{equation}
We see that we can perform a unitary transformation to
the right-handed quarks that removes all three phases $\alpha$,
$\beta$, $\gamma$. The same holds for $M_{u}$, by performing the
substitution $A \rightarrow A'$, $\Omega \rightarrow \Omega'$ and $v_i
\rightarrow v_i^*$. We note that the case I matrices were also used by Ref.~\cite{Razzaghi:2022jgq} as the mass matrices for the charged leptons.

In this work,  we study this case,  that 
corresponds to the Lagrangian in Eq.~(\ref{eq:16}). Then, given that
$D_{\alpha}D_{\alpha}^{\dagger}=\mathds{1}$ and
$D_{a}D_{a}^{\dagger}=D_{a^2} = \text{diag}(a^2,b^2,c^2)$,
we find
\begin{eqnarray}
H_d &\equiv& M_d M_d^{\dagger} = D_{v}S_d D_{v}^{\dagger}\, ,
\nonumber\\
H_u &\equiv& M_u M_u^{\dagger} = D_{v}^{\dagger}S_u D_{v}\, ,
\end{eqnarray}
where $S_d=W D_{a^2} W^{\dagger}$ and $a^2 \rightarrow a'^2$ for the
up quark case.  This matrix can now be explicitly written out using
appropriate parameters as
\begin{equation}
    S_d=
    \begin{pmatrix}
        \Sigma_d & Z_de^{i\phi_d} & Z_de^{-i\phi_d}\\ Z_de^{-i\phi_d}
        & \Sigma_d & Z_de^{i\phi_d}\\ Z_de^{i\phi_d} & Z_de^{-i\phi_d}
        & \Sigma_d
    \end{pmatrix} ,
\end{equation}
where $\Sigma_d$ and $Z_d$ are real, and
\ba
\Sigma_d &\equiv& a^2+b^2+c^2\, ,
\nonumber\\
Z_d\ e^{i\phi_d} &\equiv& a^2+\omega^2 b^2 + \omega c^2\, ,
\label{eq:Sigma_Z}
\ea
with corresponding primes for the up case.
For completeness,
the specific forms for $H_d$ and $H_u$ found after using the parameterizations
in Eqs.~\eqref{eq:vev_par} and \eqref{eq:Sigma_Z} are written
in Appendix~\ref{app:HdHu}.
The eigenvalues of the matrices $H_d$ and $H_u$ will be fitted for the
(square of the) quark masses, $(m_d^2,m_s^2,m_b^2)$ and
$(m_u^2,m_c^2,m_t^2)$, respectively

We now turn to the Cabibbo-Kobayashi-Maskawa (CKM) matrix.  As found
by Branco and Lavoura \cite{Branco:1987mj}, the absolute values of the
CKM matrix can be obtained through calculating the traces of
appropriate powers of the matrices $H_u$ and $H_d$. They observe that
\begin{equation}
  \label{eq:lavoura}
    \textrm{Tr}\left(H_u^a H_d^b\right)\equiv L_{ab}=\sum_{k,i}U_{ki}
    (D_u^a)_{kk}(D_d^b)_{ii}\, ,
\end{equation}
where $U_{ki}=|V_{ki}|^2\,$ and $V$ is the CKM matrix. The CKM matrix
is unitary and therefore $U$ only has four independent
entries. Consequently, in order to compute $U$,
it is only necessary to resort to
\begin{align}
    L_{11} =& U_{ki} (D_u)_{kk}(D_d)_{ii} \, ,
    \nonumber\\*[1mm]
    L_{12} =& U_{ki} (D_u)_{kk}(D_d^2)_{ii}\, ,
    \nonumber\\*[1mm]
    L_{21} =& U_{ki} (D_u^2)_{kk}(D_d)_{ii}\,  ,
    \nonumber\\*[1mm]
    L_{22} =& U_{ki} (D_u^2)_{kk}(D_d^2)_{ii}\, .
\end{align}
These equations are linear in $U_{ik}$ and are, therefore, invertible
for this variable.  Thus, by picking $U_{11}$, $U_{21}$, $U_{13}$, and
$U_{23}$ (respectively, $U_{ud}$, $U_{cd}$, $U_{ub}$, and $U_{cb}$),
we are able to obtain a unique solution for the magnitudes of the CKM
elements as a function of $L_{ab}$ and the quark masses.
Namely,
\begin{eqnarray}
  \label{eq:24}
    U_{11} &=& \left({m_b}^2-{m_s}^2\right)
               \left({m_c}^2-{m_t}^2\right)  \frac{
               a_{11}}{\textrm{det}}\, , 
    \nonumber\\*[1mm]
    U_{21} &=& \left({m_b}^2-{m_s}^2\right)
               \left({m_u}^2-{m_t}^2\right) \frac{
               a_{21}}{\textrm{det}}\, , 
    \nonumber\\*[1mm]
    U_{13} &=& \left({m_d}^2-{m_s}^2\right)
               \left({m_c}^2-{m_t}^2\right) \frac{
               a_{13}}{\textrm{det}}\, , 
    \nonumber\\*[1mm]
    U_{23} &=& \left({m_d}^2-{m_s}^2\right)
               \left({m_u}^2-{m_t}^2\right) \frac{
               a_{23}}{\textrm{det}}\, , 
\end{eqnarray}
where
\begin{eqnarray}
a_{11} &=&
{L_{11}} \left({m_b}^2+{m_s}^2\right)\left({m_c}^2+{m_t}^2\right)
   -{L_{12}}
   \left({m_c}^2+{m_t}^2\right)-{L_{21}}
   \left({m_b}^2+{m_s}^2\right)+{L_{22}}
\nonumber\\
&&
+m_b^2 \left(-m_c^2 m_t^2 \left(m_d^2+m_s^2\right)-m_s^2 m_u^2
   \left(m_c^2+m_t^2\right)+m_s^2 m_u^4\right)+m_c^2 m_d^2 m_t^2
   \left(m_d^2-m_s^2\right) \, , 
\\
a_{21} &=&
-{L_{11}}\left({m_b}^2+{m_s}^2\right)\left({m_t}^2+{m_u}^2\right)
+{L_{12}}\left({m_u}^2+{m_t}^2\right)+{L_{21}}\left({m_b}^2+{m_s}^2\right)-{L_{22}}
\nonumber\\
&&
+m_b^2 \left( m_c^2 m_s^2 \left(m_t^2+m_u^2-m_c^2\right)+m_t^2 m_u^2
   \left(m_d^2+m_s^2\right)\right)+m_d^2 m_t^2 m_u^2
   \left(m_s^2-m_d^2\right)\, , 
\\
a_{13} &=&
-{L_{11}}\left({m_d}^2+{m_s}^2\right)\left({m_t}^2+{m_c}^2\right)
+{L_{12}}\left({m_c}^2+{m_t}^2\right)+{L_{21}}\left({m_d}^2+{m_s}^2\right)-{L_{22}}
\nonumber\\
&&
+m_b^2 m_c^2 m_t^2 \left(m_d^2+m_s^2-m_b^2\right)+m_d^2 m_s^2 \left(m_c^2 \left(m_t^2+m_u^2\right)+m_u^2 \left(m_t^2-mu^2\right)\right) \, ,
\\
a_{23} &=&
{L_{11}}\left({m_d}^2+{m_s}^2\right)\left({m_t}^2+{m_u}^2\right)
-{L_{12}}\left({m_u}^2+{m_t}^2\right)-{L_{21}}\left({m_d}^2+{m_s}^2\right)+{L_{22}}
\nonumber\\
&&
+ m_t^2 m_u^2 \left(m_b^4-m_b^2 \left(m_d^2+m_s^2\right)-m_d^2
   m_s^2\right)+m_c^4 m_d^2 m_s^2-m_c^2 m_d^2 m_s^2
   \left(m_t^2+m_u^2\right)\, , 
\end{eqnarray}
and
\begin{equation}
\textrm{det} = \left({m_b}^2-{m_d}^2\right) \left({m_c}^2-{m_u} ^2\right)
   \left({m_d}^2-{m_s}^2\right) \left({m_u} ^2-{m_t}^2\right)\left({m_b}^2-{m_s}^2\right) \left({m_c}^2-{m_t}^2\right)\, .
\end{equation}
In these equations, the $L_{ij}$ are obtained by evaluating the
left hand side of Eq.~(\ref{eq:lavoura}). 
Finally,
we note that knowing these four CKM \textit{magnitudes},
we can determine the Jarslkog invariant \cite{Jarlskog:1985ht}, up to its sign.
Thus, given some phase convention, we are also able
to determine the phases of all CKM matrix elements.

\section{\label{sec:fitting}The fit to  the quark mass matrices}

\subsection{\label{subsec:10parameters}Parameters and observables}

We would like to fit 10 observables (6 quark masses and 4 CKM parameters) with the
10 free parameters that we have in this model,
\begin{equation}
  \label{eq:25}
  \beta_1, \beta_2, \rho_2, \rho_3, \Sigma_d, \Sigma_u, Z_d, Z_u, \phi_d,
\phi_u \, .
\end{equation}
Notice that this is a huge improvement over the SM, where there are 18 complex Yukawa parameters.
Similarly,
in Ref.~\cite{Ma:2002yp},
there are 18 Yukawa couplings; in their notation
$h_1^{u,d}$, $h_2^{u,d}$, $h_3^{u,d}$,
and those with $h \rightarrow h'$ and $h \rightarrow h''$.
These reduce to 12 complex parameters, even after the approximation in their
equation~(19).
So, having only 10 real parameters is already excellent.

Moreover,
our 10 parameters are \textit{constrained}.
Although we were not able to find an analytical relation which expresses such a
constraint,
we can show numerically that it does exist.
We postpone this proof until the end of section~\ref{subsec:results}.
The upshot is that it was not guaranteed a priori that our 10 parameters
would be able to fit the 10 observables.
Turning the argument around,
the fact that the 10 experimental values do allow for a good fit in the $A_4$-3HDM
can be viewed as a success for the model.

\subsection{\label{subsec:fit}The fitting procedure}

We have implemented a $\chi ^2$ analysis of the model, through a
minimization performed using the CERN Minuit library \cite{James:1975dr}.
The observables employed in this analysis, labeled by $i=1,...,11$ are
specified in Table~\ref{tab:obs},
where $\overline{X}_i$ represents the experimental mean value of the
observable $X_i$ and $\sigma_i$ is the experimental error, which, when
both left and right bounds are stated, is assumed to be the largest of
the two.
\begin{table}[h]
\begin{center}
\begin{tabular}{ c|c|c } 
 Observable & Experimental value & Model prediction \\ \hline
 $m_u$ [MeV] & $2.16 \pm 0.50$ & $2.15$ \\ 
  $m_c$ [MeV] & $1270 \pm 20$ & $1271.6$  \\ 
  $m_t$ [GeV] & $172.69 \pm 0.30$ & $172.68$  \\ 
  $m_d$ [MeV] & $4.67 \pm 0.50$ & $4.66$ \\ 
  $m_s$ [MeV] & $93.4\pm 8.6$ & $92.08$  \\ 
  $m_b$ [MeV] & $4180\pm 30$ & $4180.39$  \\ 
  $|V_{11}|$ & $0.97435\pm0.00016$ & $0.97434$ \\ 
  $|V_{21}|$ & $0.22486\pm0.00067$ & $0.22479$ \\ 
  $|V_{13}|$ & $0.00369 \pm0.00011$ & $0.00369$  \\ 
  $|V_{23}|$ & $0.04182\pm0.00085$ &  $0.04178$ \\ 
  $J$ & $(3.08\pm0.15)\times 10^{-5} $ &  $3.09\times 10^{-5}$ \\ 
 \hline
\end{tabular}
\caption{\label{tab:obs}Experimental values and fit results.}
\end{center}
\end{table}
The data on the quark masses as well as for the CKM matrix
elements and the Jarlskog invariant experimental values 
were obtained from \cite{ParticleDataGroup:2022pth}.
As mentioned,
$|J|$ is fixed by  $|V_{11}|$, $|V_{21}|$, $|V_{13}|$, and $|V_{23}|$.
However,
using it in the fit speeds the numerical convergence onto a good solution.

The $\chi ^2$ function depends on the 10 parameters of our model \eqref{eq:25},
\begin{equation}
  \label{eq:25}
  \beta_1, \beta_2, \rho_2, \rho_3, \Sigma_d, \Sigma_u, Z_d, Z_u, \phi_d,
\phi_u
\end{equation}
and is written as 
\begin{equation}
    \chi^2(\textrm{\textbf{p}})= \sum_{i=1}^{11} \left(
      \frac{P_i(\textrm{\textbf{p}})-\overline{X}_i}{\sigma_i}
    \right)^2 \, , 
\end{equation}
where $P_i(\textrm{\textbf{p}})$ is our model's prediction for each of
the 11 (10 + $J$) observables. The fit is complicated by the fact that the masses
(squared) are obtained from the eigenvalues of $H_d,H_u$ but the
elements of the CKM also depend on the masses, see
Eq.~(\ref{eq:24}). So, we start by calculating the eigenvalues of
$H_d$ and $H_u$, which depend only on the parameters in Eq.~(\ref{eq:25}).
Then,
we evaluate the $L_{ij}$ from the left hand side of
Eq.~(\ref{eq:lavoura}), and finally the CKM elements are obtained from
Eq.~(\ref{eq:24}). In Appendix~\ref{app:HdHu}
 we give the explicit expressions for the matrices $H_d$ and $H_u$.

\subsection{\label{subsec:results}Results of the fit}

We have found an excellent fit of our model to the data,
given in the
second column of Table \ref{tab:obs}. This fit results in $\chi^2=
0.058$, for the parameters 
\begin{align}
\beta_1 =&1.4260868\ \textrm{radians} \, ,
\nonumber\\
\beta_2 =& 1.5424328\ \textrm{radians}  \, ,
\nonumber\\
\rho_2 =&4.2784971\  \textrm{radians}  \, ,
\nonumber\\
\rho_3 =&5.3682785\  \textrm{radians}  \, ,
\nonumber\\
\Sigma_d =& 0.2889178\times 10^{-3} \, ,
\nonumber\\
\Sigma_u =&0.4927455 \, ,
\nonumber\\
Z_d =&0.1816577\times 10^{-3} \, ,
\nonumber\\
Z_u =&0.4758317 \, ,
\nonumber\\
\phi_d =&-1.7324779 \ \textrm{radians}  \, ,
\nonumber\\
\phi_u=&0.20644967\times 10^{-1} \ \textrm{radians} \, .
\end{align}
\\
This fit also leads to the data in the third column of Table
\ref{tab:obs}, as well as to the vevs 
\begin{equation}
    |v_i| =
\left(1.00604, 6.90357 , 245.901\right) \textrm{(GeV)}.
\label{eq:vev_numerical}
\end{equation}
We notice that the vevs obey $v_1<v_2<<v_3$. 
This hierarchy of vevs is related to the hierarchy of the quark
masses. This was also obtained in Ref.~\cite{Lavoura:2007dw}, although
their model is not consistent, as their vev structure is not that of
\cite{Degee:2012sk} for the symmetric $A_4$ potential they consider.

We can now perform a second (toy) fitting procedure,
which illustrates the fact that the ten parameters in our model are
\textit{constrained},
as announced at the end of section~\ref{subsec:10parameters}.
In this fit, we take all experimental values in Table~\ref{tab:obs},
\textit{except} that we trade the correct experimental value
of $m_s$ for $ m_s=(2 \pm 0.02) \textrm{GeV}$.
Now, the fit is very poor, having $\chi^2=600$.
If these had been the correct experimental values for the 10 observables,
then our model would not be able to fit them.
Conversely,
the fact that such a fit is possible is a success for the model.

\section{\label{sec:viable}Viability of the vacuum found in the fit}

We start by defining the three
doublets as in Eq.~(\ref{eq:2}). 
Next we define the physical eigenstates for the charged Higgs as
$(G^+,S^+_2,S^2_3)^T$, and for the neutral states we have
$(G^0,S^0_2,S^0_3,S^0_4,S^0_5,S^0_6)^T$, identifying the would-be Goldstone bosons
$G^+\equiv S^+_1$ and $G^0\equiv S^0_1$.
With these conventions,
and  following the definitions in \cite{Grimus:2007if},
we define the $3\times 3$ matrix $\tilde{U}$ by
\begin{equation}
  \label{eq:3}
  \varphi_i^+ \equiv \sum_{j=1}^3 \tilde{U}_{ij} S^+_j\,,
\end{equation}
and the $3\times 6$ matrix $\tilde{V}$ by
\begin{equation}
  \label{eq:4}
  x_i + i x_{i+3} = \sum_{j=1}^6 \tilde{V}_{ij} S^0_j\,.
\end{equation}
These matrices\footnote{From the point of view of a simultaneous
fit of the Yukawa and scalar sectors,
it is a pity that these matrices $\tilde{V}$ and $\tilde{U}$ have in the literature
the same notation as the CKM matrix $V$ and $U_{ki} = |V_{ki}|^2$.}
are then related to the diagonalization matrices of the
charged and neutral scalars, to which we now turn.

\subsection{\label{sec:min}
The minimization of the potential}

In our procedure we already know the values of the vevs. So, we use
the stationarity equations to solve for the soft parameters, and leave
the quartic parameters of the potential $\Lambda_i$ as free
parameters. In this way we can solve for $m^2_{11}, m^2_{22}, m^2_{33}$ as
well as for $\text{Im} (m^2_{12}), \text{Im} (m^2_{13})$,
leaving as free parameters the $\Lambda_i$ and
$\text{Re} (m^2_{12}), \text{Re} (m^2_{13}), \text{Re} (m^2_{23}),\text{Im}
(m^2_{23})$. When evaluating the scalar mass matrices (see below) the conditions
have to be applied to ensure that we are at the minimum. For
completeness we write these conditions in Appendix~\ref{app:min_cond}.

\subsection{The charged mass matrix}

The charged mass matrix is obtained from the second derivatives at the
minimum,
\begin{equation}
  \label{eq:5}
  \mathcal{M}^2_C = \left.\frac{\partial^2 V_H}{\partial
    \varphi_i^+\partial\varphi_j^-}\right|_{\text{Min}} .
\end{equation}
The matrix $\mathcal{M}^2_C$ is an hermitian matrix, with real
eigenvalues and satisfying, with our usual conventions,
\begin{equation}
  \label{eq:6}
  R_{\rm ch}  \mathcal{M}^2_C  R^\dagger_{\rm ch}
  =\text{diag}(0,m^2_{S^+_2},m^2_{S^+_3})\equiv \mathcal{M}^2_{D_{ch}}\,,
\end{equation}
where $R_{\rm ch}$ is an unitary matrix that satisfies,
\begin{equation}
  \label{eq:7}
  S^+_i = \sum_{j=1}^3 \left(R_{\rm ch}\right)_{ij} \varphi^+_j\,.
\end{equation}
This can be seen from
\begin{align}
  \label{eq:8}
  \mathcal{L}_{\rm mass}=& - \varphi^-_i
  \left(\mathcal{M}^2_C\right)_{ij}\varphi^+_j =
  - \varphi^-_i
  \left(R^\dagger_{\rm ch} R_{\rm ch} \mathcal{M}^2_C R^\dagger_{\rm ch} R_{\rm ch}
  \right)_{ij} \varphi^+_j
  = - \varphi^-_i
  \left(R^\dagger_{\rm ch} \mathcal{M}^2_{D_{ch}} R_{\rm ch}
  \right)_{ij} \varphi^+_j\nonumber\\
  =&- S^-_i \left(\mathcal{M}^2_{D_{ch}}\right)_{ij} S^+_j\,,
\end{align}
where we have used Eq.~(\ref{eq:7}).

We have checked both algebraically
and numerically that we have a zero eigenvalue corresponding to $G^+$
and we require that all other masses squared are positive, a condition
for a local minimum.

\subsection{The neutral mass matrix}

Since in our case CP is not conserved, we denote the unrotated neutral
scalars by $x_i, i=1,\ldots,6$, as in Eq.~(\ref{eq:2}). We therefore
obtain the neutral mass matrix as,
\begin{equation}
  \label{neutral_mass}
    \mathcal{M}^2_N = \left.\frac{\partial^2 V_H}{\partial
    x_i\partial x_j}\right|_{\text{Min}} .
\end{equation}
This is a symmetric real matrix diagonalized by an orthogonal
$6\times 6$ matrix,
\begin{equation}
  \label{eq:9}
R_{\rm neu}  \mathcal{M}^2_N  R^T_{\rm neu}
=\text{diag}(0,m^2_{S^0_2},m^2_{S^0_3},m^2_{S^0_4},m^2_{S^0_5},m^2_{S^0_6})
\equiv \mathcal{M}^2_{D_{\rm neu}}\,,
\end{equation}
with
\begin{equation}
  \label{eq:12}
  S^0_i = \sum_{j=1}^6 \left(R_{\rm neu}\right)_{ij} x_j\,.
\end{equation}
As for the case of the charged scalars, we have checked both algebraically
and numerically that we have a zero eigenvalue corresponding to $G^0$
and we require that all other masses squared are positive, a condition
for a local minimum.

\section{\label{sec:theorconstraints}Theoretical Constraints}

After having shown that a solution exists for the vevs and parameters
in the Yukawa sector that correctly fits the quarks masses and the CKM
entries, we have to show that this is compatible with the scalar
potential analysis. In particular we have to show that the vevs
correspond to a local minimum of the potential and that both the
theoretical constraints as well as those coming from LHC are satisfied.
In this section we analyze the theoretical constraints.

\subsection{Perturbative Unitarity}

This problem was already solved in \cite{Bento:2022vsb}, so we take
the potential in the form of Eq.~(\ref{V4_A4_one}). From
Ref.~\cite{Bento:2022vsb} we have the following expression for the
eigenvalues $\lambda_i$\footnote{We use $\lambda_i$ instead of
  $\Lambda_i$, in order to not confuse with the notation of Eq.~(\ref{V_A4}).}
\begin{align}
      \lambda_1=&2 \left(2 \text{Re}(c_3)+r_1\right)\\
      \lambda_2=&2 \left(\sqrt{3}\, |\text{Im}(c_3)| -\text{Re}(c_3) +
        r_1\right)\\ 
      \lambda_3=&2 \left(-\sqrt{3}\, |\text{Im}(c_3)| -\text{Re}(c_3)
        + r_1\right)\\
      \lambda_4=&2 (r_4+r_7)\\
      \lambda_5=&2 (r_4-r_7)\\
      \lambda_6=&2 (r_1+2 r_7)\\
      \lambda_7=&2 (r_1-r_7)\\
      \lambda_8=&2 (r_4+|c_3|)\\
      \lambda_9=&2 (r_4-|c_3|)\\
      \lambda_{10}=&6 r_1+8 r_4+4 r_7\\
      \lambda_{11}=&6 r_1-2 (2 r_4+r_7)\\
      \lambda_{12}=&6 |c_3|+2 r_4+4 r_7\\
      \lambda_{13}=&-6 |c_3|+2 r_4+4 r_7
\end{align}
Perturbative unitarity is satisfied if
\begin{equation}
  \label{pert_unit}
  |\lambda_i| < 8 \pi, \quad \forall i .
\end{equation}

\subsection{The BFB conditions}

For the $A_4$ symmetric  potential, the conditions for boundedness from below along the neutral directions
(BFB-n) have been conjectured in \cite{Ivanov:2020jra}, and proved to
hold in \cite{Buskin:2021eig}. These are
\begin{align}
&
\Lambda_0 + \Lambda_3 \geq 0\, ,
\label{BFB_1b}
\\
&
\frac{4}{3}(\Lambda_0 + \Lambda_3) + \frac{1}{2} (\Lambda_1 + \Lambda_2)
- \Lambda_3
- \frac{1}{2} \sqrt{(\Lambda_1 - \Lambda_2)^2+\Lambda_4^2} \geq 0\, ,
\label{BFB_2b}
\\
&
\Lambda_0 + \frac{1}{2} (\Lambda_1 + \Lambda_2)
+ \frac{1}{2} (\Lambda_1 - \Lambda_2)
\cos{(2 k \pi/3 )}
+ \frac{1}{2} \Lambda_4
\sin{(2 k \pi/3 )} \geq 0\, \ \ (k=1,2,3)\, .
\label{BFB_3b}
\end{align}

However, as shown in \cite{Faro:2019vcd,Ivanov:2020jra},
a potential which is BFB-n is not necessarily BFB along the charge
breaking directions (BFB-c). 
Necessary BFB-c conditions have yet to be found for the $A_4$ 3HDM,
but sufficient conditions have been proposed in
\cite{Carrolo:2022oyg} 
following the technique developed in \cite{Boto:2022uwv}. They are,
\begin{equation}
A_d \geq 0\, ,
\ \ \ \ 
A_o \geq -A_d/2\, ,
\label{copositivity2b}
\end{equation}
where
\begin{align}
A_d =& a\ =\ \frac{2}{3}(\Lambda_0+\Lambda_3)\, ,
\nonumber\\[+2mm]
A_o =& b + \textrm{min}(0,c)-d
\nonumber\\
=&
\frac{1}{3}(2\Lambda_0 - \Lambda_3)
+\frac{1}{2}(\Lambda_1+\Lambda_2) +
\min(0,-\frac{1}{2}(\Lambda_1+\Lambda_2)) -
\frac{1}{2} \sqrt{(\Lambda_1-\Lambda_2)^2+\Lambda_4^2}\, .
\label{Ad_Ao}
\end{align}

It is important to remark that, since these are sufficient, but not
necessary, conditions, some good points in parameter space may be
excluded by this restriction.

\subsection{The oblique parameters $S,T,U$}

For this we use the notation and results from
\cite{Grimus:2007if}, which require the matrices $\tilde{U}$ and $\tilde{V}$.
Comparing Eq.~(\ref{eq:7}) with the definition in
Eq.~(\ref{eq:3}), we conclude that
\begin{equation}
  \label{eq:10}
  \tilde{U} = R_{\rm ch}^\dagger \,,
\end{equation}
where the matrix $R_{\rm ch}$ is obtained from the numerical
diagonalization of Eq.~(\ref{eq:6}).
Similarly,
comparing Eq.~(\ref{eq:12}) with the definition of $\tilde{V}$ in
Eq.~(\ref{eq:4}), we get,
\begin{equation}
  \label{eq:13}
  \tilde{V}_{ij}=
    \left(R^T_{\rm neu}\right)_{ij}+i\left(R^T_{\rm
        neu}\right)_{i+3,j} .
\end{equation}
Having $\tilde{U}$ and $\tilde{V}$, we can construct the needed matrices
$\text{Im}\left(\tilde{V}^\dagger \tilde{V} \right)$ ,
$\tilde{U}^\dagger \tilde{U}$,
$\tilde{V}^\dagger \tilde{V}$
and $\tilde{U}^\dagger \tilde{V}$, and implement the procedure of \cite{Grimus:2007if}.

\subsection{Global minimum}

After finding a set of $m_{i,j}$ and $\Lambda_i$ which reproduce the
vevs in Eq.~\eqref{eq:vev_numerical} necessary for a good fit of the
quark mass matrices, and after performing the previous theoretical
checks on the scalar potential, we must still ensure that our minimum
is indeed the global minimum.  This step is almost never taken in
studies of quark mass matrices, since there are no exact analytical
formulae for it.  Moreover, one must check that there are no lower
minima both along the neutral directions and along the charge breaking
directions.  We follow the strategy discussed in
Ref.~\cite{Carrolo:2022oyg}.
Take a specific set of $m^2_{ij}$ and
$\Lambda_i$. Then we parameterize the scalar doublets as~\cite{Faro:2019vcd,Carrolo:2022oyg},
\begin{equation}
  \label{eq:26}
\langle \phi_1 \rangle
=\sqrt{r_1}\,
  \begin{pmatrix}
    0\\*[1mm]
    1
  \end{pmatrix}\, ,
\quad
\langle \phi_2 \rangle
=\sqrt{r_2}\,
  \begin{pmatrix}
    \sin(\alpha_2)\\*[1mm]
    \cos(\alpha_2) e^{i \beta_2}
  \end{pmatrix}\, ,
\quad
\langle \phi_3 \rangle
=\sqrt{r_3} e^{i \gamma}\,
  \begin{pmatrix}
    \sin(\alpha_3)\\*[1mm]
    \cos(\alpha_3) e^{i \beta_3}
  \end{pmatrix}\, ,
\end{equation}
where we have already used the gauge freedom.
Now we let the vevs run free, for both charge conserving and charge
violating directions. We give one seed point and perform a minimization
of the potential using the CERN Minuit library~\cite{James:1975dr}.
We obtain not only the value of the potential at the minimum, but also
the values of $r_i,\alpha_2,\beta_2,\alpha_3,\beta_3$ and $\gamma$.
Then, we take one more (randomly generated) seed point and repeat the
minimization.
Finally, we take the minimum as the global one if it is found as
the global minimum in each of 200 searches with randomly generated
seed points. We have done this verification for every point that
passed all the constraints. In all cases, we found that the local
minimum was also a global minimum. In particular we always found that
\begin{equation}
  \label{eq:27}
  \sin(\alpha_2)=\sin(\alpha_3)=0,
\end{equation}
showing that we do not have charged breaking
directions\footnote{
  To cross check our numerical procedure we also
  considered points that violated the BFB conditions. And, indeed for
  these points, our algorithm showed that the potential was not BFB and could
  have charge breaking directions as well.}
and, comparing
with Eq.~(\ref{eq:2}), we verified numerically that,
\begin{equation}
  \label{eq:28}
 \frac{|v_i|}{\sqrt{2}}= \sqrt{r_i},\quad
  e^{i\, \rho_2}= \cos(\alpha_2)\, e^{i\, \beta_2},\quad
  e^{i\, \rho_3}= \cos(\alpha_3)\, e^{i\, (\beta_3+\gamma)}\, .
\end{equation}

\section{\label{sec:LHC}Simple LHC Constraints}

Up to now we have implemented the theoretical constraints on the
model. The next step is to implement the LHC constraints. To do this
completely one would have to implement all the decays of the neutral and
charged Higgs as well as their branching ratios.
One would also have to worry about the 
electric dipole moments (EDM) and the flavour-changing
neutral couplings (FCNC), as the model does not have a structure of
couplings of the Higgs to the fermions that automatically ensures
vanishing FCNC \cite{Glashow:1976nt,Ferreira:2010xe,Yagyu:2016whx}.
This lies beyond the scope of the present work.
Nonetheless, we can implement easily the constraints 
that come from $h \rightarrow WW/ZZ$ in the $\kappa$ formalism,
where the deviation from the coupling of the SM Higgs boson
to a pair of $W$'s (or $Z$'s) is measured by $\kappa_V$.
In our model,
\begin{equation}
  \label{k_V}
  \kappa_V=R^{\rm neu}_{21} v_1 +  R^{\rm neu}_{22} v_2 \cos(\rho_2) +
  R^{\rm neu}_{23} v_3 \cos(\rho_3) +  R^{\rm neu}_{25} v_2 \sin(\rho_2) +
  R^{\rm neu}_{26} v_3 \sin(\rho_3),
\end{equation}
where $R^{\rm neu}$ is matrix defined in Eq.~(\ref{eq:9}).
We take the experimental 
constraint from ATLAS~\cite{ATLAS:2022vkf},
\begin{equation}
  \label{eq:29}
  \kappa_W= 1.0206{\,}^{+ 0.05172}_{- 0.05087},\quad
  \kappa_Z= 0.99{\,}^{+ 0.06136}_{- 0.05214}\, .
\end{equation} 

\section{\label{sec:results}Results}

In this section we present the results of the analysis of the scalar
potential after imposing that we have a good solution for the fit of
the quarks masses and CKM entries, as explained in
Section~\ref{sec:fitting}. 

\subsection{Scanning strategy}

We start by imposing the vevs obtained in the fit.
\begin{equation}
  \label{eq:20}
  v_1=1.00604\, \text{(GeV)},\quad v_2=6.90357\, e^{i\, 4.278497}\, \text{(GeV)},
  \quad v_3=245.901\, e^{i\, 5.368278} \, \text{(GeV)} .
\end{equation}
Now we vary the free parameters of the potential in the following ranges,
\begin{equation}
  \label{eq:21}
  \log_{10}|\Lambda_i| \in[-3,1],
\quad  \log_{10}|\text{Im}(m^2_{23})|  \in [-1,7] \text{GeV}^2,
\quad \log_{10}|\text{Re}(m^2_{ij})| \in [-1,7] \text{GeV}^2,
\end{equation}
where in the last equation we use
\begin{equation}
  \label{eq:23}
   m^2_{ij} \in \left\{m^2_{12},m^2_{13},m^2_{23} \right\} .
\end{equation}
We randomly scan as in Eq.~(\ref{eq:21}), and then:
\begin{enumerate}
\item Apply the theoretical constraints that only depend on the
  $\Lambda_i$, that is BFB and perturbative unitarity.
\item Then obtain the eigenvalues for the charged and neutral
  scalars. Verify that all the masses squared are positive, and that
  we have a zero eigenvalue corresponding to the Goldstone bosons,
  $G^0$ and $G^+$. 
\item Verify the S, T and U oblique parameters.
\item Apply the LHC constraint on $\kappa_V$.
\item Check numerically that the vev is indeed a global minimum.

\end{enumerate}

\subsection{The scalar spectrum}
\label{sec:spectrum}

We found that there is a strong correlation in the scalar masses.
If we denote the masses
of the neutral scalars by
$(m_{G^0}=0,m_{S^0_{2}},m_{S^0_{3}},m_{S^0_{4}},m_{S^0_{5}},m_{S^0_{6}})$, 
and $(m_{G^+}=0,m_{H^+_{1}},m_{H^+_{2}})$ for the charged scalars, we
find numerically that
\begin{equation}
  \label{eq:15}
  m_{S^0_{3}}\simeq m_{S^0_{4}}\simeq m_{H^+_{1}},\quad
  m_{S^0_{5}}\simeq m_{S^0_{6}}\simeq m_{H^+_{2}} .
\end{equation}
This is true even if we do not require $m_{S^0_{2}}=125$ GeV, and
specially true after implementing the constraints of perturbative unitarity, BFB
and STU. But, as we want to reproduce the LHC results, we also required
that~\cite{ParticleDataGroup:2022pth}
\begin{equation}
  \label{eq:19}
  m_{S_2^0}= 125.25 \pm 0.17\quad \text{GeV}.
\end{equation}

In the following figures we show the correlation among the masses.
Included in red are the points generated before the theoretical cuts
were applied, 
and in green the points remaining after the constraints were implemented.

\begin{figure}[H]
  \centering
  \begin{tabular}{cc}
    \includegraphics[width=0.48\textwidth]{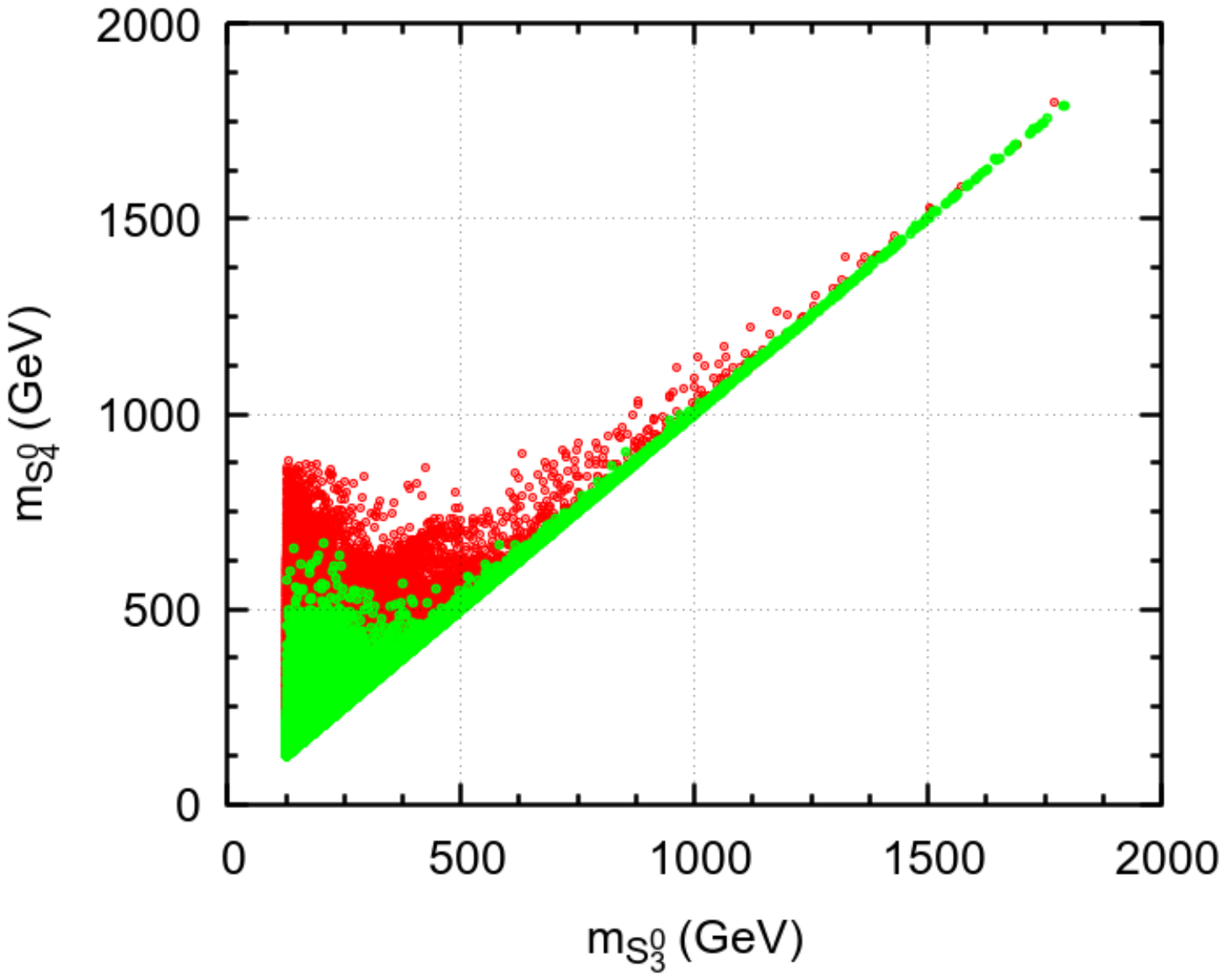}&
    \includegraphics[width=0.48\textwidth]{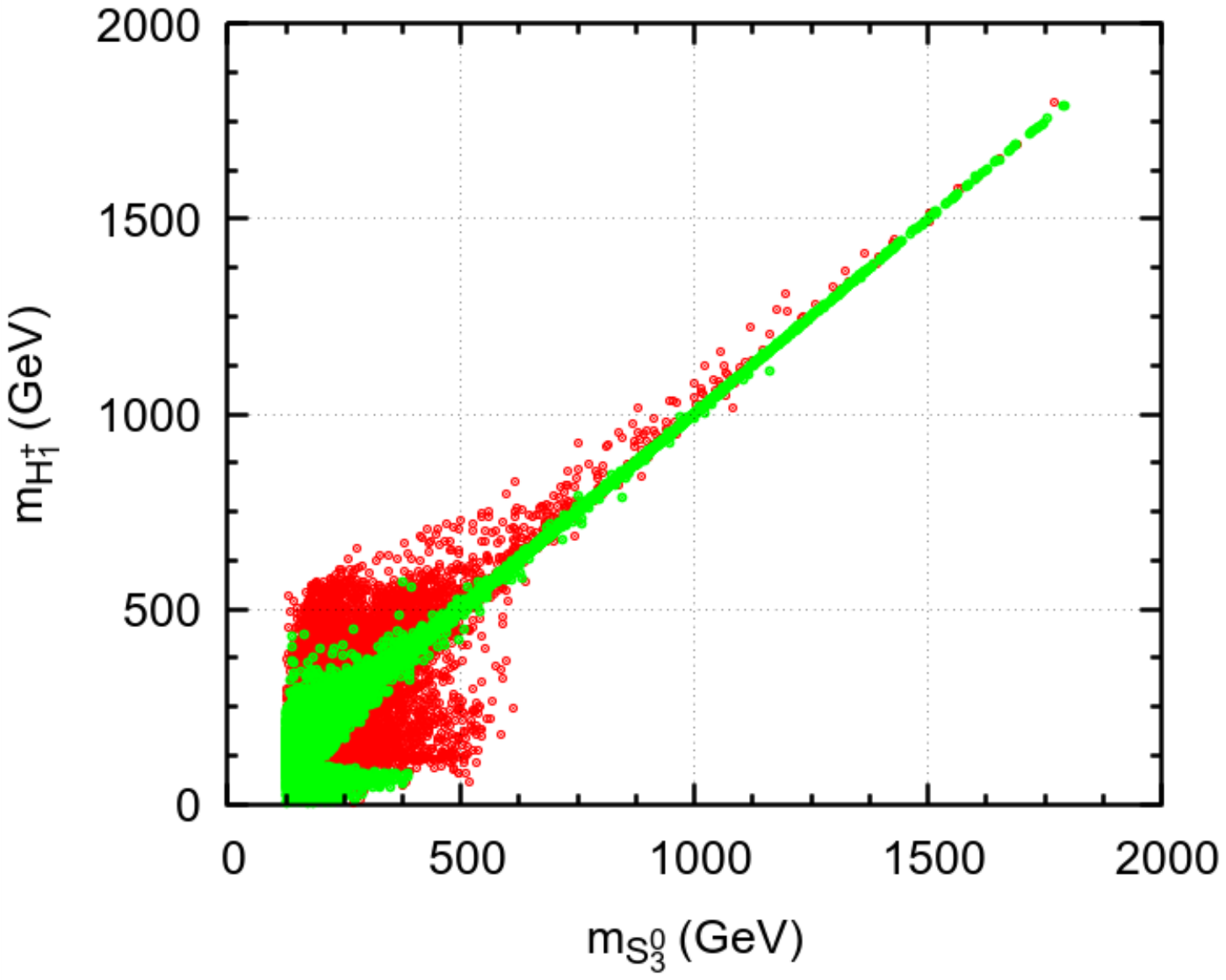}
  \end{tabular}
  \caption{Left panel: Relation between $m_{S_3^0}$ and
    $m_{S_4^0}$. Right panel: Relation between $m_{S_3^0}$ and
    $m_{H_1^+}$. Color conventions: No cuts (red); with cuts (green)}
  \label{fig:1}
\end{figure}
\begin{figure}[H]
  \centering
  \begin{tabular}{cc}
    \includegraphics[width=0.48\textwidth]{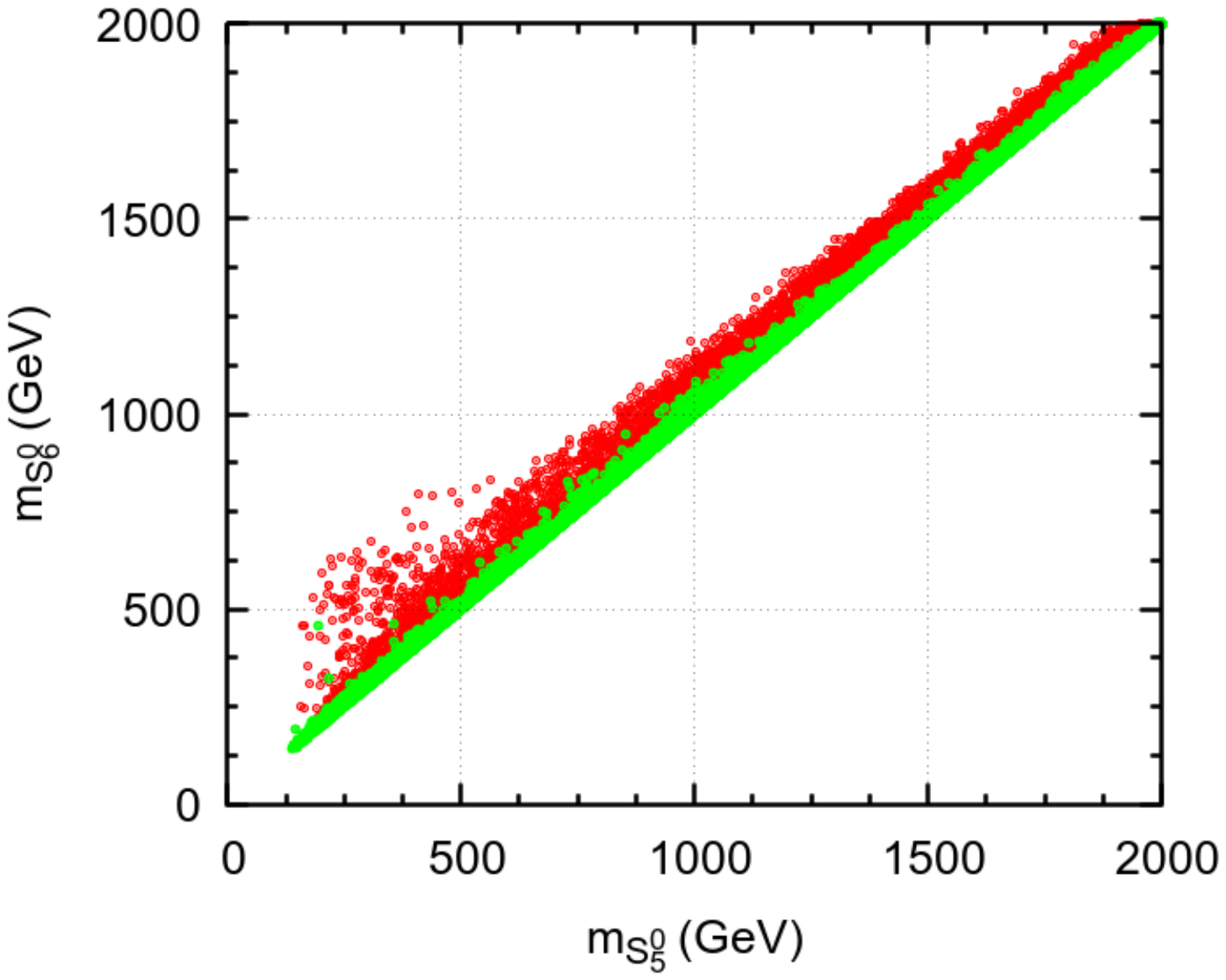}&
    \includegraphics[width=0.48\textwidth]{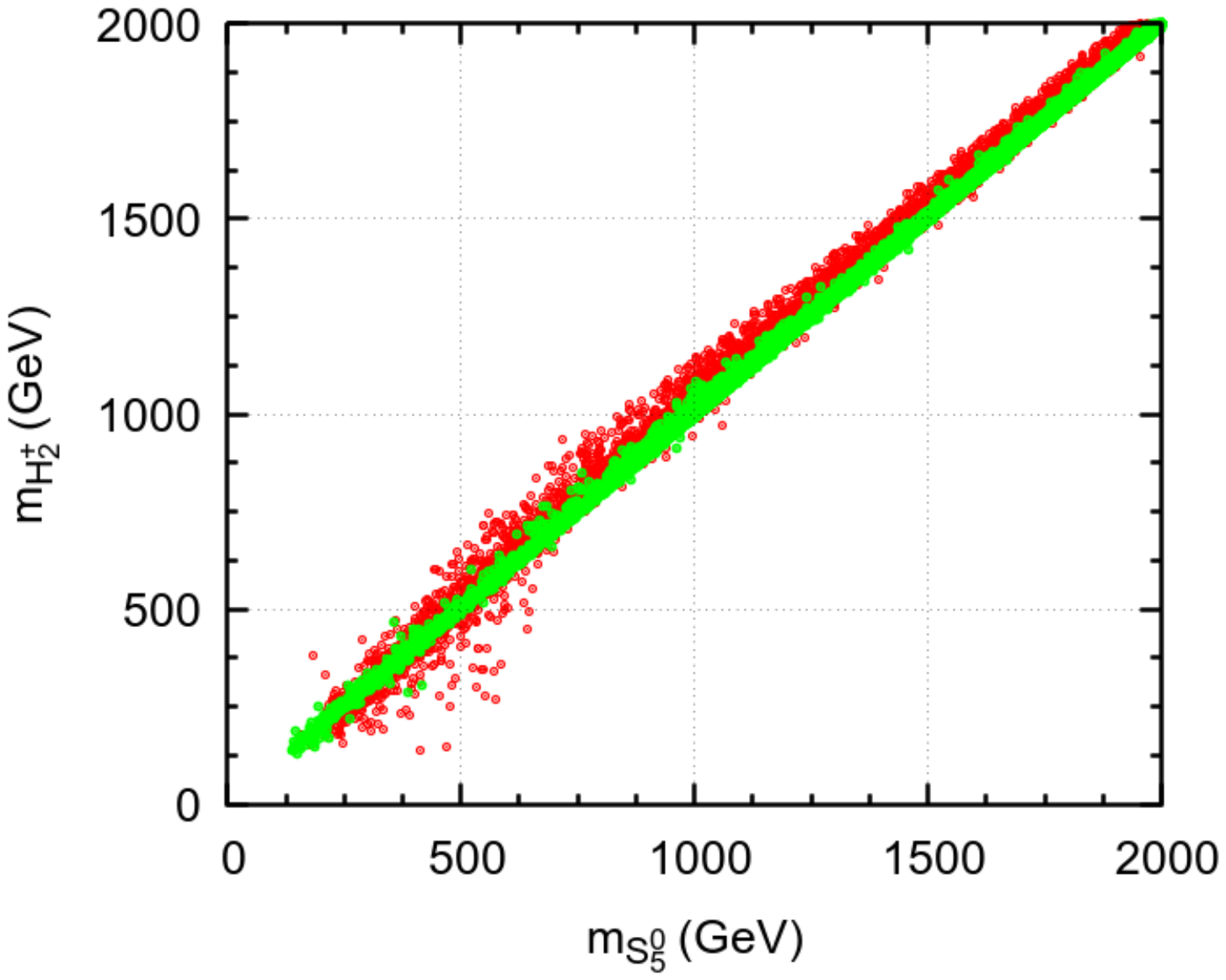}
  \end{tabular}
  \caption{Left panel: Relation between $m_{S_5^0}$ and
    $m_{S_6^0}$. Right panel: Relation between $m_{S_5^0}$ and
    $m_{H_2^+}$. Color conventions: No cuts (red); with cuts (green)}
  \label{fig:2}
\end{figure}
\begin{figure}[H]
  \centering
  \begin{tabular}{cc}
    \includegraphics[width=0.48\textwidth]{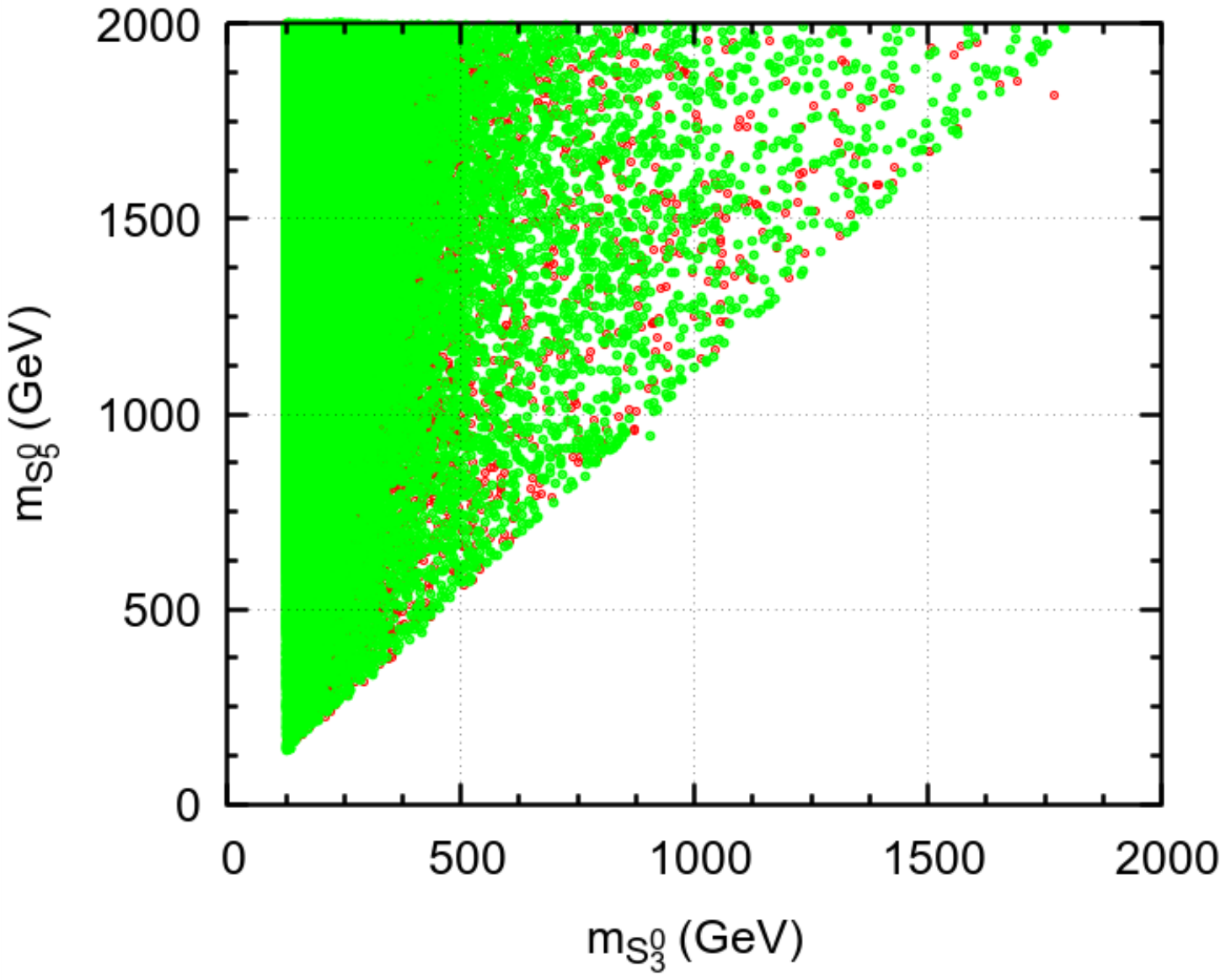}&
    \includegraphics[width=0.48\textwidth]{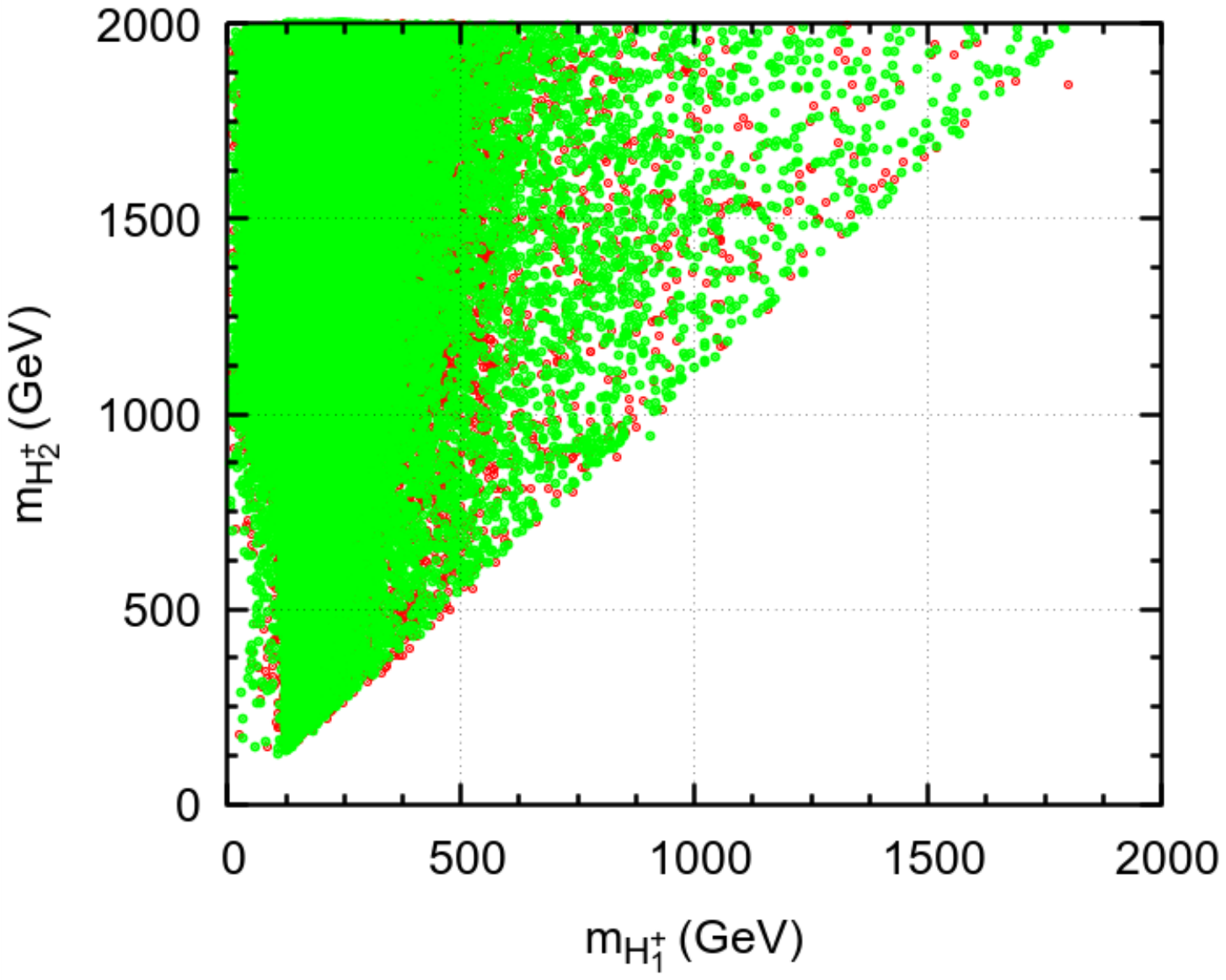}
  \end{tabular}
  \caption{Left panel: Relation between $m_{S_3^0}$ and
    $m_{S_5^0}$. Right panel: Relation between $m_{H_1^+}$ and
    $m_{H_2^+}$. Color conventions: No cuts (red); with cuts (green)}
  \label{fig:3}
\end{figure}

\subsection{The $\kappa_V$ constraint}

\begin{figure}[H]
  \centering
  \begin{tabular}{cc}
    \includegraphics[width=0.48\textwidth]{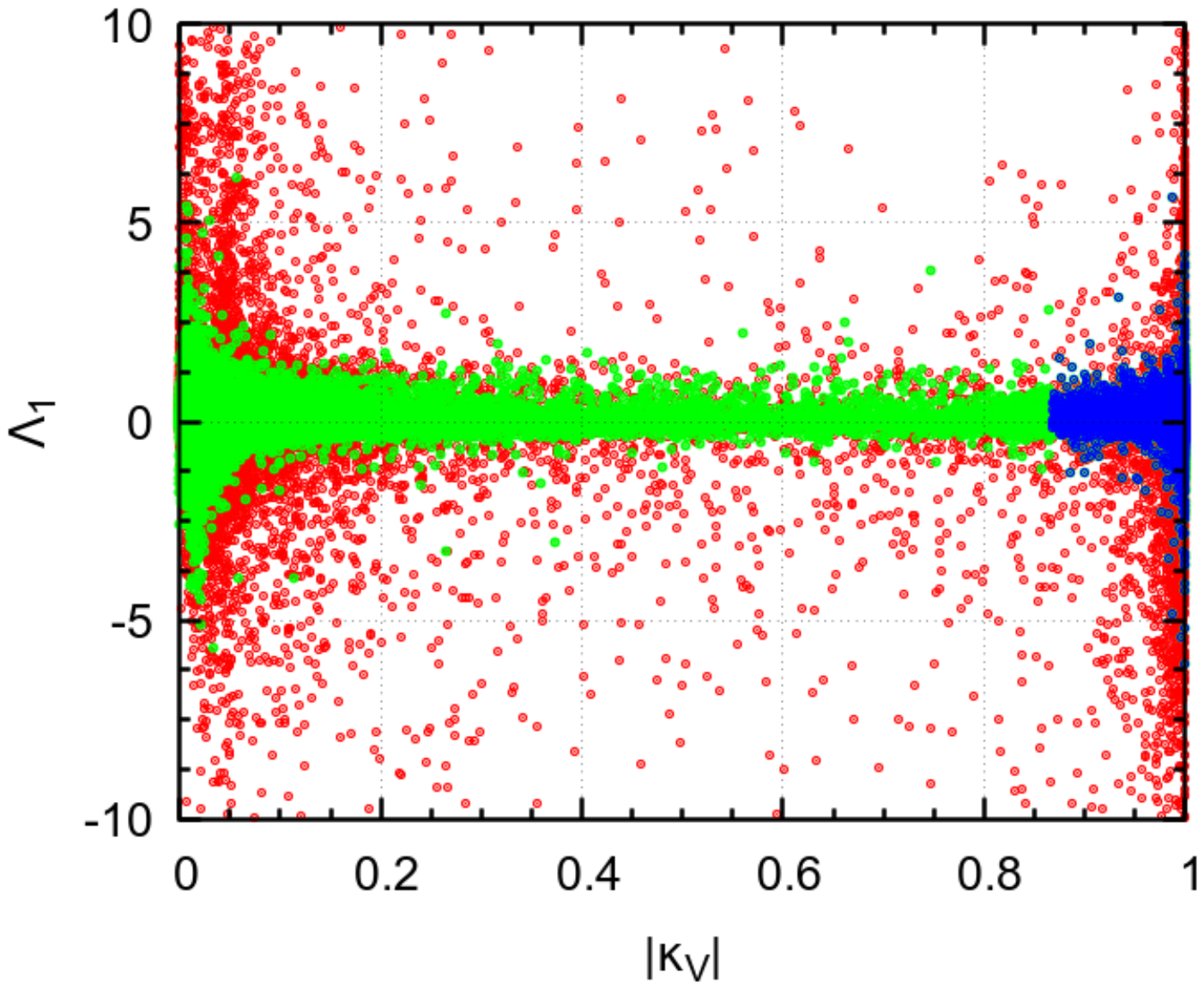}&
    \includegraphics[width=0.48\textwidth]{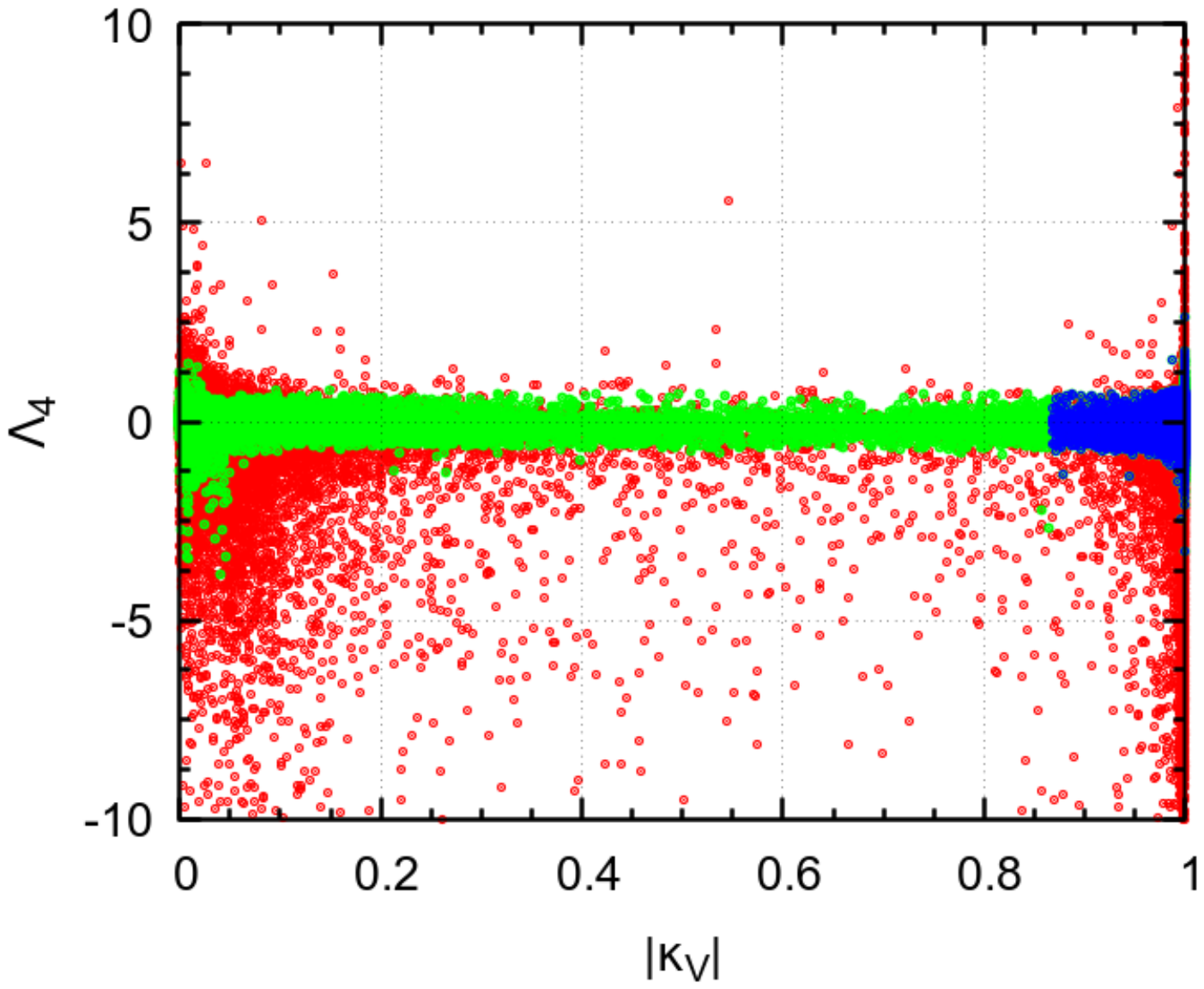}
  \end{tabular}
  \caption{Left panel: Relation between $\kappa_V$ and
    $\Lambda_1$. Right panel: Relation between $\kappa_V$ and
    $\Lambda_4$. Color
    conventions: No cuts (red), with theoretical cuts (green), and
    after the $\kappa_V$ constraint (blue).} 
  \label{fig:4}
\end{figure}

\begin{figure}[H]
  \centering
  \begin{tabular}{cc}
    \includegraphics[width=0.48\textwidth]{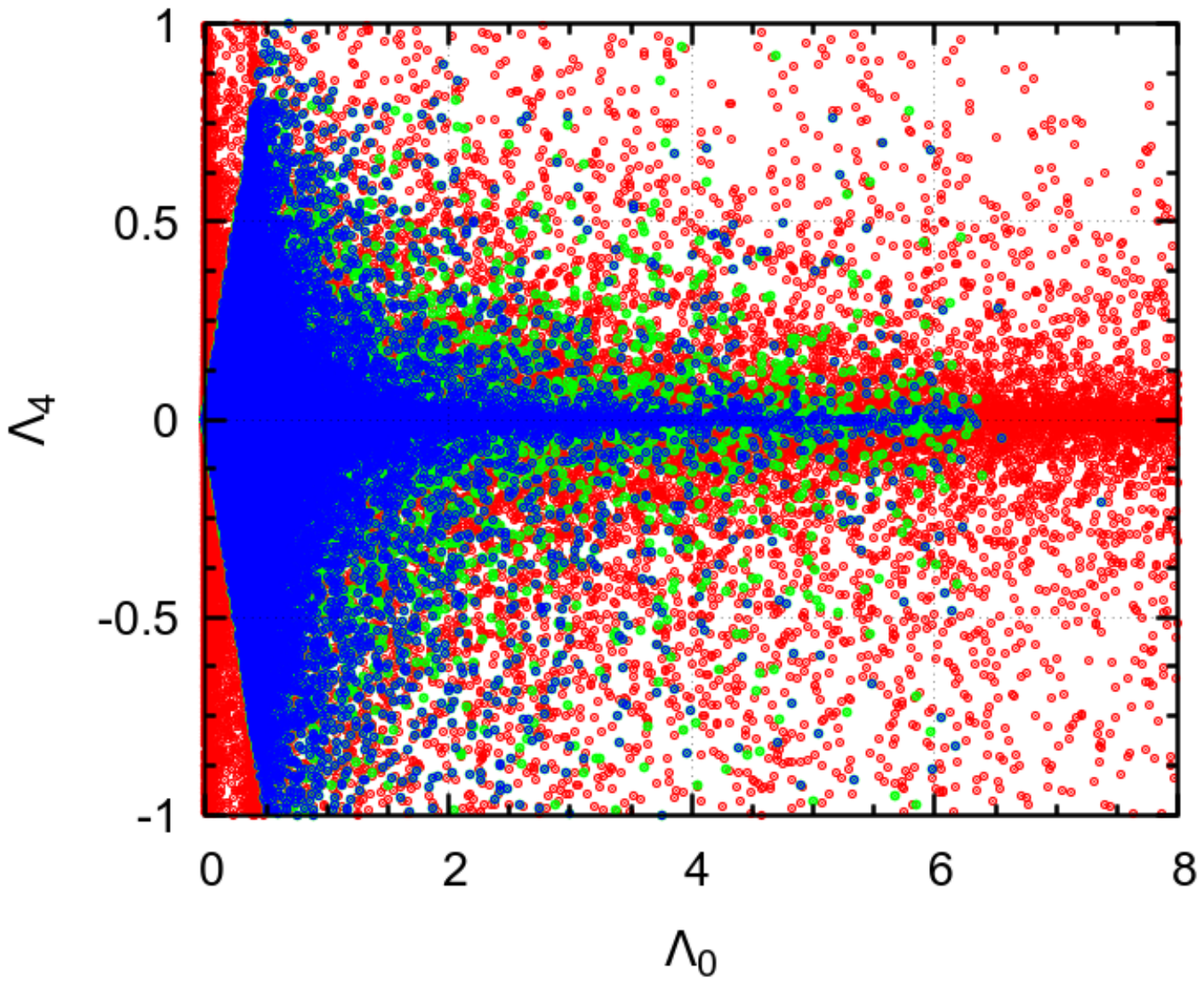}&
    \includegraphics[width=0.48\textwidth]{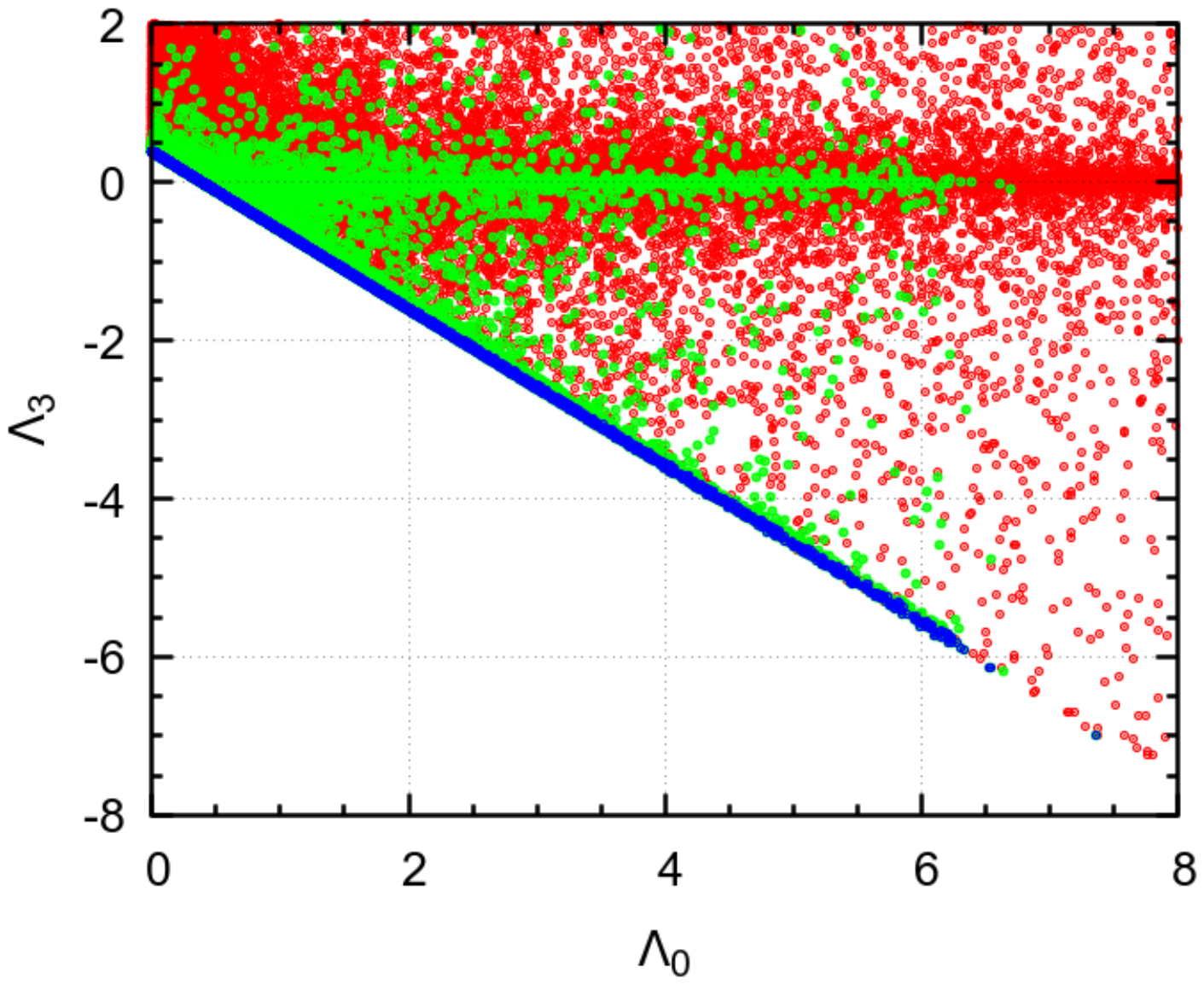}
  \end{tabular}
  \caption{Left panel: Relation between $\Lambda_0$ and
    $\Lambda_4$. Right panel: Relation between $\Lambda_0$ and
    $\Lambda_3$ Color
    conventions: No cuts  (red), with theoretical cuts (green), and
    after the $\kappa_V$ constraint (blue).} 
  \label{fig:5}
\end{figure}

We can now implement the $\kappa_V$ constraint on the model.
In the following figures,
in red are points without cuts, in green with cuts but no
$\kappa_V$ constraint, and finally in blue points remaining after
this constraint is applied.
We took the ATLAS result of Eq.~(\ref{eq:29}) at $2\sigma$.
While the theoretical constraints cut around 88\% of the
points, the $\kappa_V$ constraint only cuts 22\% of the remaining points.
In Fig.~\ref{fig:4} we show the relation between $\kappa_V$ and
$\Lambda_{1,4}$ for the three sets of points as discussed above.
In fact it is not obvious from Fig.~\ref{fig:4} that the $\kappa_V$
constraint only cuts about 22\% of the points that pass the other
cuts. This is because there is
a very large number of points with $|\kappa_V|\lesssim 1$, even without
theoretical cuts, and this is even more so after imposing the theoretical cuts.
In this figure, we have 200000 points in the green region,
but from these 156516 are in the blue region.
That is, after theoretical cuts, 78\% of the
points also satisfy the  $\kappa_V$ constraint.
In Fig.~\ref{fig:5} we show the relation between $\Lambda_0$ and
$\Lambda_{3,4}$ for the same sets of points. 
\noindent
We see that, while for $(\Lambda_0,\, \Lambda_4)$ there is not much difference
before and after the $\kappa_V$ constraint, the same is not true for
$(\Lambda_0,\, \Lambda_3)$, where the constraints impose a linear relation
between those two parameters. We note that, while $\Lambda_0$ is
always positive, $\Lambda_3$ can be negative respecting the BFB
condition in Eq.~(\ref{BFB_1b}), $\Lambda_0+\Lambda_3 \geq 0$, as it
is clear in the right-handed panel of Fig.~\ref{fig:5}.
Before we end this section, let us remark that we did not redraw the
figures in Sec.~\ref{sec:spectrum} after imposing the $\kappa_V$
constraint, as the blue points would just superimpose the green
points, as we have checked.

\section{\label{sec:concl}Conclusions}

It is known that the 3HDM symmetric under
an exact $A_4$ symmetry is not compatible with non-zero quark masses
and/or non-block-diagonal CKM matrix \cite{GonzalezFelipe:2013xok}.
In this work, we studied a 3HDM with $A_4$ softly broken.
This allows us to evade the above result, by enlarging the structure
of the possible vacua. 

We obtained an excellent fit of the quarks mass matrices, including
the CP-violating Jarlskog invariant.
This leads to a unique solution for the vevs. 
We showed that, with the solution for the vevs obtained from the fit, it
is possible to have a local minimum of the potential. We enforce this
by imposing that all squared masses are positive. As in our scheme the scalar
masses are not input parameters, we have to restrict one of the neutral
scalars to have the mass of the known Higgs boson.

We have implemented the BFB, perturbative unitarity and the
oblique parameters $S,T,U$ theoretical constraints. 
From LHC, we have considered the observed Higgs mass and the $\kappa_V$
constraint.\footnote{The detailed study of other LHC constraints as well
as those coming from FCNC and the EDM lies
beyond the scope of the present work, and is left for a future publication.}
After imposing the other constraints, we found that most of the points are
close to the alignment required to respect the experimental $\kappa_V$ constraint.
We have discovered a strong correlation among the masses of the
scalars, even before applying the theoretical constraints, especially
for moderate to large scalar masses.

One important point is that we have numerically checked for all the
points that pass our constraints, that for a given set of
parameters of the potential, our minimum is the true global minimum.

\vspace{5ex}

\newpage

{\Large \textbf{Acknowledgments}}

\noindent
This work is supported in part by FCT (Fundação para a Ciência e
Tecnologia) under Contracts
CERN/FIS-PAR/0002/2021,
CERN/FIS-PAR/0008/2019,
UIDB/00777/2020,
and UIDP/00777/2020;
these projects are partially funded through POCTI (FEDER), COMPETE,
QREN, and the EU. The work of I.B. was supported by a CFTP fellowship
with reference BL210/2022-IST-ID and the work of S. C. by a CFTP
fellowship with reference BL255/2022-IST-ID.

\appendix

\section{\label{app:HdHu}The matrices $H_d$ and $H_u$}

\begin{align}
  \label{eq:Hd}
      H_d(1,1)=&\Sigma_d v^2 \cos^2(\beta_1) \cos^2(\beta_2)\\
      H_d(1,2)=&  v^2 Z_d \cos(\beta_1) \cos^2(\beta_2)
      \cos(\rho_2 - \phi_d)
      \sin(\beta_1) \nonumber\\
      &
      - i\, v^2 Z_d \cos(\beta_1) \cos^2(\beta_2)
        \sin(\beta_1) \sin(\rho_2 - \phi_d)\\
        H_d(1,3)=&  v^2 Z_d \cos(\beta_1) \cos(\beta_2)
        \cos(\rho_3 + \phi_d)
        \sin(\beta_2)\nonumber\\
        &
        - i\, v^2 Z_d \cos(\beta_1) \cos(\beta_2) 
          \sin(\beta_2) \sin(\rho_3 + \phi_d)\\
      H_d(2,1)=&(H_d(1,2))^*\\
      H_d(2,2)=&\Sigma_d v^2 \cos^2(\beta_2) \sin^2(\beta_1)\\
      H_d(2,3)=&   v^2 Z_d \cos(\beta_2)
      \cos(\rho_2 - \rho_3 + \phi_d) \sin(\beta_1) 
        \sin(\beta_2)\nonumber\\
      & + i\, v^2 Z_d \cos(\beta_2) \sin(\beta_1) 
        \sin(\beta_2) \sin(\rho_2 - \rho_3 + \phi_d)\\
      H_d(3,1)=&(H_d(1,3))^*\\
      H_d(3,2)=&(H_d(2,3))^*\\
      H_d(3,3)=&\Sigma_d v^2 \sin^2(\beta_2)
\end{align}

\begin{align}
  \label{eq:Hu}
      H_u(1,1)=&\Sigma_u v^2 \cos^2(\beta_1) \cos^2(\beta_2)  \\    
      H_u(1,2)=&  v^2 Z_u \cos(\beta_1) \cos^2(\beta_2)
      \cos(\rho_2 + \phi_u)
        \sin(\beta_1) \nonumber\\
      &+ i\, v^2 Z_u \cos(\beta_1) \cos^2(\beta_2)
        \sin(\beta_1) \sin(\rho_2 + \phi_u)\\
        H_u(1,3)=&  v^2 Z_u \cos(\beta_1) \cos(\beta_2)
        \cos(-\rho_3 + \phi_u)
        \sin(\beta_2) \nonumber\\
      &- i\, v^2 Z_u \cos(\beta_1) \cos(\beta_2)
          \sin(\beta_2)\sin(-\rho_3 + \phi_u)\\
      H_u(2,1)=&(H_u(1,2))^*\\
      H_u(2,2)=&\Sigma_u v^2 \cos^2(\beta_2) \sin^2(\beta_1)\\
      H_u(2,3)=&   v^2 Z_u \cos(\beta_2)
      \cos(-\rho_2 + \rho_3 + \phi_u) \sin(\beta_1)
        \sin(\beta_2) \nonumber\\
      &+ i\, v^2 Z_u \cos(\beta_2) \sin(\beta_1)
        \sin(\beta_2) \sin(-\rho_2 + \rho_3 + \phi_u)\\
      H_u(3,1)=&(H_u(1,3))^*\\
      H_u(3,2)=&(H_u(2,3))^*\\
      H_u(3,3)=&\Sigma_u v^2 \sin^2(\beta_2)
\end{align}

\section{\label{app:min_cond}The minimization conditions}

\begin{align}
m^2_{11}=&
-\frac{\sec (\rho_2) \sec (\rho_3)}{24 v_1^2}
 \left[-12 \text{Im}(m^2_{23}) v_2
   v_3 \sin (2 (\rho_2-\rho_3))+\cos (\rho_2-\rho_3) \left(4
     \Lambda_0 v_1^2 v^2\right.\right.
   \nonumber\\
   &\hskip 10mm
   \left.\left.
     +6
   \Lambda_1 v_1^2 v_2^2+6 \Lambda_1 v_1^2 v_3^2+3
   \Lambda_1 v_2^2 v_3^2-3 \Lambda_2 v_2^2 v_3^2+2
   \Lambda_3 v_1^2 \left(2
     v_1^2-v_2^2-v_3^2\right)\right)\right.
   \nonumber\\
   &\hskip 10mm
\left.
 +4 \Lambda_0 v_1^2
   v_2^2 \cos (\rho_2+\rho_3)+4 \Lambda_0 v_1^2 v_3^2 \cos
   (\rho_2+\rho_3)+4 \Lambda_0 v_1^4 \cos (\rho_2+\rho_3)\right.
   \nonumber\\
   &\hskip 10mm
\left.
   +6
   \Lambda_1 v_1^2 v_2^2 \cos (\rho_2+\rho_3)
   +6 \Lambda_1
   v_1^2 v_3^2 \cos (\rho_2+\rho_3)-3 \Lambda_1 v_2^2
   v_3^2 \cos (3 (\rho_2-\rho_3))\right.
   \nonumber\\
   &\hskip 10mm
   \left.
     +3 \Lambda_2 v_2^2 v_3^2
   \cos (3 (\rho_2-\rho_3))
   -2 \Lambda_3 v_1^2 v_2^2 \cos
   (\rho_2+\rho_3)-2 \Lambda_3 v_1^2 v_3^2 \cos
   (\rho_2+\rho_3)\right.
   \nonumber\\
   &\hskip 10mm
   \left.
     +4 \Lambda_3 v_1^4 \cos (\rho_2+\rho_3)+3
   \Lambda_4 v_1^2 v_2^2 \sin (\rho_2-\rho_3)+3 \Lambda_4
   v_1^2 v_2^2 \sin (\rho_2+\rho_3)\right.
   \nonumber\\
   &\hskip 10mm
   \left.
     +3 \Lambda_4 v_1^2
   v_3^2 \sin (\rho_2-\rho_3)-3 \Lambda_4 v_1^2 v_3^2 \sin
   (\rho_2+\rho_3)-3 \Lambda_4 v_2^2 v_3^2 \sin
   (\rho_2-\rho_3)\right.
   \nonumber\\
   &\hskip 10mm
   \left.
     +3 \Lambda_4 v_2^2 v_3^2 \sin (3
   (\rho_2-\rho_3))-12 \text{Re}(m^2_{23}) v_2 v_3 \cos (2
   (\rho_2-\rho_3))\right.
   \nonumber\\
   &\hskip 10mm
   \left.
     +24 \text{Re}(m^2_{13}) v_1 v_3 \cos (\rho_2)+24
   \text{Re}(m^2_{12}) v_1 v_2 \cos (\rho_3)+12 \text{Re}(m^2_{23}) v_2
   v_3\right]
\end{align}

 \begin{align}
   m^2_{22}=&
   -\frac{1}{12 v_2}
   \Big[3 \sec (\rho_2) \left(-4 \text{Im}(m^2_{23}) v_3 \sin
   (\rho_3)+v_2 v_3^2 (\Lambda_1-\Lambda_2) \cos (\rho_2-2
   \rho_3)-\Lambda_4 v_2 v_3^2 \sin (\rho_2-2 \rho_3)\right.
\nonumber\\
&\hskip 10mm
\left.
   +4
   \text{Re}(m^2_{23}) v_3 \cos (\rho_3)+4 \text{Re}(m^2_{12})
   v_1\right)+v_2 \left(4 \Lambda_0
   v^2 +6 \Lambda_1 v_1^2+3
   \Lambda_1 v_3^2+3 \Lambda_2 v_3^2\right.
\nonumber\\
&\hskip 10mm
\left.
  -2 \Lambda_3 v_1^2+4
   \Lambda_3 v_2^2
  -2 \Lambda_3 v_3^2+3 \Lambda_4 v_1^2 \tan
   (\rho_2)\right)\Big]
 \end{align}

 \begin{align}
   m^2_{33}=&
   -\frac{1}{12 v_3}
   \Big[3 \sec (\rho_3) \left(4 \text{Im}(m^2_{23}) v_2 \sin
   (\rho_2)+v_2^2 v_3 (\Lambda_1-\Lambda_2) \cos (2
   \rho_2-\rho_3)-\Lambda_4 v_2^2 v_3 \sin (2
   \rho_2-\rho_3)\right.
\nonumber\\
&\hskip 10mm
\left.
   +4 \text{Re}(m^2_{23}) v_2 \cos (\rho_2)+4 \text{Re}(m^2_{13})
   v_1\right)+v_3 \left(4 \Lambda_0
   \left(v_1^2+v_2^2+v_3^2\right)+6 \Lambda_1 v_1^2+3
   \Lambda_1 v_2^2\right.
\nonumber\\
&\hskip 10mm
\left.
  +3 \Lambda_2 v_2^2-2 \Lambda_3 v_1^2-2
   \Lambda_3 v_2^2+4 \Lambda_3 v_3^2-3 \Lambda_4 v_1^2 \tan
   (\rho_3)\right)\Big]
 \end{align}

 \begin{align}
   \text{Im}(m^2_{12})=&
   \frac{1}{4 v_1}
   \Big[\sec (\rho_2) \left(4 \text{Im}(m^2_{23}) v_3 \cos
   (\rho_2-\rho_3)-\Lambda_1 v_2 v_3^2 \sin (2
   (\rho_2-\rho_3))-\Lambda_1 v_1^2 v_2 \sin (2
   \rho_2)\right.
\nonumber\\
&\hskip 5mm
\left.
  +\Lambda_2 v_2 v_3^2 \sin (2
   (\rho_2-\rho_3))+\Lambda_2 v_1^2 v_2 \sin (2
   \rho_2)-\Lambda_4 v_2 v_3^2 \cos (2
   (\rho_2-\rho_3))\right.
\nonumber\\
&\hskip 5mm
\left.
  +\Lambda_4 v_1^2 v_2 \cos (2 \rho_2)-4
   \text{Re}(m^2_{23}) v_3 \sin (\rho_2-\rho_3)-4 \text{Re}(m^2_{12}) v_1
   \sin (\rho_2)\right)\Big]
 \end{align}

 \begin{align}
   \text{Im}(m^2_{13})=&
   -\frac{1}{4 v_1}
   \Big[\sec (\rho_3) \left(4 \text{Im}(m^2_{23}) v_2 \cos
   (\rho_2-\rho_3)-\Lambda_1 v_2^2 v_3 \sin (2
   (\rho_2-\rho_3))+\Lambda_1 v_1^2 v_3 \sin (2
   \rho_3)\right.
\nonumber\\
&\hskip 5mm
\left.
  +\Lambda_2 v_2^2 v_3 \sin (2
   (\rho_2-\rho_3))-\Lambda_2 v_1^2 v_3 \sin (2
   \rho_3)-\Lambda_4 v_2^2 v_3 \cos (2
   (\rho_2-\rho_3))\right.
\nonumber\\
&\hskip 5mm
\left.
  +\Lambda_4 v_1^2 v_3 \cos (2 \rho_3)-4
   \text{Re}(m^2_{23}) v_2 \sin (\rho_2-\rho_3)+4 \text{Re}(m^2_{13}) v_1
   \sin (\rho_3)\right)\Big]
\end{align}

\providecommand{\href}[2]{#2}\begingroup\raggedright\endgroup


\end{document}